\documentclass[12pt,a4paper,]{article}
\usepackage{amsmath,amssymb,amsfonts,amsthm}

\usepackage[latin1]{inputenc}
\usepackage[english]{babel}
\usepackage{graphics}
\usepackage{amsfonts}

\newcommand\bd{\begin{displaymath}}
\newcommand\ed{\end{displaymath}}
\newcommand\vt{\vert}
\newcommand\vv{\vert^2}

\begin{document}

\title{Finite temperature corrections and embedded strings in noncommutative geometry and the
standard model with neutrino mixing}

\author{R. A. Martins\footnote{Email: rmartins@math.ist.utl.pt, research supported by Fundaç\~ao para a Ci\^encias e a Tecnologia (FCT) including program POCI 2010/FEDER.} \\\\ Centro de An\'alise Matem\'atica,\\ Geometria e Sistemas Din\^amicos \\ Departamento de Matem\'atica \\ Instituto Superior T\'ecnico 
\\ Av. Rovisco Pais \\ 1049-001 Lisboa}
\date{\today}
\maketitle

\begin{abstract}
The recent extension of the standard model to include massive neutrinos in  the framework of
noncommutative geometry and the spectral action principle involves new scalar fields and their
interactions with the usual complex scalar doublet. After ensuring that they bring no unphysical consequences, we address the question of how these fields
affect the physics predicted in Weinberg-Salam theory, particularly in the context of
the Electroweak phase transition. Applying the Dolan-Jackiw procedure, we calculate the finite temperature corrections, and find that the phase transition is first order. The new scalar interactions significantly improve the stability of the Electroweak $Z$ string, through the ``bag'' phenomenon described by Watkins and Vachaspati. (Recently cosmic strings have climbed back into interest due to new evidence). Sourced by static embedded strings, an internal space analogy of Cartan's torsion is drawn, and a possible Higgs-force-like `gravitational' effect of this non-propagating torsion on the fermion masses is described. We also check that the field generating the Majorana mass for the $\nu_R$ is non-zero in the physical vacuum. \end{abstract}

\newpage

\tableofcontents

\section{Introduction}
The noncommutative geometric formulation of the standard model coupled to  gravity by Connes and
Chamseddine (the spectral action principle \cite{sap}) has been extended to include massive neutrinos
\cite{smv}\footnote{The validity of the Majorana type terms in found in this article should be considered modulo the fact that the calculations were done in the Euclidean signature.}, \cite{lng}, \cite{snm} \cite{gsn}.

By definition, the standard  model describes 3 massless chirality left neutrinos $\nu_L$ and
precludes chirality right neutrinos $\nu_R$. Further to the results of laboratory experiments it is
nowadays consensual that neutrinos are massive particles and due to lack of evidence, the question
of the existence of $\nu_R$ is open. Theoretical studies of particle physics now include
appropriate generalisations of the standard model, but as the theory of Weinberg, Glashow and Salam
has been shown to be very strong in its predictive power, the modifications are made as minimally as
possible.

In particular, the generalisation  referred to above is done by the
omission of the unaxiomatic assumption of $S^o$-reality and the natural inclusion of $\nu_R$. However, it cannot be said that nothing else in the model may alter as a consequence; the most notable new feature being the appearance of additional scalar fields. These are note-worthy because such a change can radically alter a theory: for example, it is normally considered undesirable to involve new fields, as they are likely to contradict experiment.

Indeed in  the first studies of such new fields \cite{leptoquarks} called leptoquarks, they cause colour symmetry breaking, which is of course absolutely unacceptable. More positively, in the extension being
discussed here, some of the new fields are colour-charge-free, and might be as difficult to detect
as $\nu_R$, while the leptoquarks are either non-dynamical, or ruled out altogether by an additional constraint applied to the Dirac operator \cite{gsn}.

To add  scalar fields to the theory is by no means bound to be a trivial alteration and given that
the extension does not contradict experiment, let us instead ask the affirmative question of what
predictions it may lead to.

Recently there has been a lot of fresh interest in cosmic strings since two things have changed over
the last few years. Cosmic strings are expected because of their occurrence in many supersymmetry grand unified theories, but due to inconsistencies between predictions and the cosmic microwave background temperature anisotropies, it was necessary to explain why they and other topological defects should exist but do not. The favourite explanation is that they were ``inflated away'' beyond observable horizon. Topological defects were once thought to provide the seeds of structure formation, but these data ruled them out. The two new things that have happened \cite{RCS} are (a) with recent advancements in theoretical physics, cosmic superstrings are expected to play the role of cosmic strings, and (b) more importantly, the data that ruled them out is being revised. Recent evidence implies that the parameters may be constrained sufficiently to allow agreement between predictions and measurements. 

At the level of Weinberg-Salam theory, there are no topologically stable defects as the vacuum manifold, the three-sphere, is simply connected. Nevertheless solitonic solutions there are, which are referred to as embedded strings and sphalerons. These embedded defects although classically stable, have been shown to be unstable to small perturbations within a realistic range of parameters. Watkins and Vachaspati \cite{VW} realised that an extension to the standard model by additional scalar fields forming bound states with the Electroweak $Z$ string can improve its stability. It is clearly appropriate to ask whether the new fields in the extension we are studying can contribute to the stability in a similar vein. What we find is that the equations of motion precisely describe the scenario predicted in \cite{VW}.

The following questions are motivated:

\begin{itemize}
\item What is the precise field content of the extended theory, and does this generalisation have
unphysical consequences? 
\item Is the mass generating field for the $\nu$ non-zero in the physical vacuum?
\item What effect if any do the new scalar fields have on the order of the
Electroweak phase transition? In other  words, do they ensure a gradual or a discrete change in the
phase?
\item Do they have any stabilising effect on the embedded strings?
\item What is the role played by these fields in the context of internal space physics as described by Connes in \cite{gravity and matter}? That is, in the equivalence principle where the Higgs takes on the role of the gravitational field strength as a connection on the 0 dimensional internal space.
\end{itemize}

In the first section we establish the total scalar field content of the theory, explaining why
there is no conflict with experiment, and show that the new scalar that generates the Majorana mass
for $\nu_R$ is non-zero in the physical vacuum. We also discuss a Wick a rotation from the Euclidean signature to the Lorentzian for the case of the model with $KO$-dimension 0. Secondly, we introduce the Coleman-Weinberg \cite{cw}
mechanism for calculating the quantum corrections at zero temperature in $\phi^4$-theory, and then
work through the details in the Dolan-Jackiw calculation of the finite temperature
corrections \cite{dj} in the same theory. The latter procedure is applied to the said noncommutative geometry
extension of the standard model, and the Ehrenfest classification is used to determine the order
of the Electroweak phase transition, analysing the effect of the new fields on this physical
process. Finally, we show that the new fields can greatly increase the stability of Electroweak $Z$ strings and discuss an internal space version of torsion gravity sourced by the strings and its effect on the fermion masses in addition to that of the Higgs force. Kibble and Lepora \cite{KL} have studied two metrical structures on the Electroweak vacuum manifold. Applying their idea, we find a set of scalar metrics, one for each fermion mass.

The calculation to determine the order of phase transition is carried out for \emph{both} the
Euclidean and the Lorentzian signatures, and both signatures are studied
below in section \ref{field content}. Now we explain how the positive features in each lead us to consider them each in
turn.

The extension in question was first done in the Euclidean signature \cite{smv} and Majorana-type terms
were identified, but the physical meaning of these terms was unclear. In \cite{snm} and \cite{lng} a modification
of the axioms was made, and a way to carry out the extension in the Lorentzian signature was
realised with the KO-dimension of the spectral triple 6 instead of 0. This time, there was a genuine Majorana mass term for the $\nu_R$ whose mass was generated via a
see-saw-like mechanism, by a new scalar field separate from the Higgs. Moreover, the
fermion quadrupling problem was also solved. Thus a completely valid physical theory was developed,
in particular, as the Gelfand-Naimark theorem has no counterpart to non-compact nor noncommutative
spaces.

Now we give the reasons why we will nevertheless, also include the Euclidean signature case.
Firstly, the internal space Einstein's equations of \cite{smv} led to a fully degenerate set of
eigenvalues for the fermion mass matrix. However, with the further constraint that leptoquarks be defined to be non-dynamical, the
correct fermion mass matrix turned out to be a solution of the field equations: the solutions had features compatible with experiment even in addition to the non-degeneracy of the masses, and for this reason it is arguably unlikely that these results came out correctly by coincidence \cite{tsm}. Further to this result, we will carry out our calculations in the Euclidean and consider the
Wick rotation to the Lorentzian signature to take place at the end. The reason why these solutions were not included in \cite{smv} was because the aim there was to consider the result of varying all the degrees of freedom in the finite triple Dirac operator, in other words no parameters were to be held constant. At first then, it appeared unnatural that some fields be non-dynamical. Also, the most general form for the Dirac operator was found mathematically instead of the mass matrix being added in by hand. Here, we avoid colour symmetry breaking not by adding in a new constraint, but by providing a reason why the leptoquarks might be constants. In the final section of this paper we explain that the static strings with improved stability can source non-propagating internal space torsion, and hence they provide a physical reason why the leptoquarks should forfeit their equations of motion. 

Moreover, from a mathematical point of view,  a `noncommutative space' is the generalisation of a Riemannian manifold to noncommutative geometry rather than that of a pseudo-Riemannian manifold, and therefore, in doing calculations in the Euclidean
signature, nothing of the original spirit of noncommutative geometry is lost; a minimal
generalistion of the Gelfand-Naimark theorem to an `almost commutative geometry' would maintain local
compactness, while the representation of the algebra would still act on a Hilbert space, which by
definition has a positive definite signature. In addition, it has been revealed that there is no Hochschild cycle in the Lorentzian model with right-handed neutrinos that can generate the chirality \cite{noa}. 

The latter is a mathematical argument only, and since the mathematics is not complete for the $KO$-dimension 6 almost-commutative theory (by the arguments above for example) anyway, we argue that further study of massive right-handed neutrinos in the Lorentzian signature in the current context should not be ruled out: we might expect that a counterpart of the Gelfand-Naimark theorem to Lorentzian almost-commutative geometries would lead to equivalent results as those of the $KO$-dimension 0 model after a Wick rotation.

In order for this article to make coherent sense, it is necessary to give a minimal amount of
background information on the articles we cite. Details on the subject of the spectral action
principle in noncommutative geometry and previous work are kept to a minimum. In contrast, we  will elaborate a little more on the physics of phase transitions in order to motivate the calculations that follow, and also work through Dolan and Jackiw's derivation in more detail than is already available in the literature.

\subsection{Background information on noncommutative geometry}

Connes' noncommutative geometry is built by the notion of a spectral triple defined as follows, and
the spectral action principle is the theory of the standard model coupled to gravity whose action
is derived by calculating the eigenvalues of the Dirac operator that acts on a tensor product space
of the spectral triple of space-time and the 0 dimensional internal space, finite spectral triple.
(Internal space is similar to the extra dimensions in Kaluza-Klein theory; it represents the
charges on the particles.) As a diffeomorphism invariant geometrical theory, the spectral action
principle provides an important step towards the unification of particle physics with gravity.
Moreover, the Higgs takes on the geometrical role of connection on internal space, and hence is the
field strength in the internal space equivalent of gravity.

A spectral  triple $(\mathcal{A},\mathcal{H},\mathcal{D})$ provides the analogue of a Riemannian
spin manifold to noncommutative geometry. It consists of $\textbf{1.}$ a real, involutive, not
necessarily commutative algebra $\mathcal{A}$, $\textbf{2.}$ A Hilbert space $\mathcal{H}$ on which
the algebra is represented and $\textbf{3.}$ A Dirac operator $\mathcal{D}$ that gives a notion of
distance, and from which is built a differential algebra. The geometry of any closed Riemannian
spin manifold can be fully described by a spectral triple and a non-commutative geometry is
essentially the same structure but with the generalisation that the algebra of coordinates are
allowed to be non-commuting. In the case of the standard model 0 dimensional internal space
spectral triple, the Hilbert space represents the fermions, the Dirac operator is the fermion mass
matrix and the algebra is derived from the symmetries of the action such that the latter be the
internal automorphisms (lifted to the spinors) of the algebra.

The Dirac  operator encodes the metric, and Sch\"ucker et al (for one generation of elementary
fermions) calculated Einstein's equations for internal space by minimising the Higgs potential -
this is the part of the spectral action that depends on the internal space Dirac operator only, also called the finite spectral action - with respect to the fluctuations of the Dirac
operator \cite{Schucker}. They found that the standard model fermion mass matrix was a solution. In \cite{smv} the
equations were calculated using a different method: all the fluctuations of the Dirac operator were
varied, including the Yukawa couplings and Cabbibo-Kobayashi-Maskawa angles, thus promoting these
traditional constants to dynamical variables. This was done for two reasons: to apply the idea that
in gravity all the degrees of freedom in the metric should be varied, and to attempt to remove the
need for choice in the pre-fluctuation Dirac operator. To this second type of Einstein equation,
the solution is that the fermion masses are all the same, but it does identify which masses are
non-zero and which are zero.

The standard model spectral triple is a product of two spectral triples, one representing (Euclidean) space-time and the other representing the internal space of particle charges. The latter has a finite dimensional Hilbert space and a noncommutative algebra. The tensor product algebra is `almost commutative':

\begin{equation}
   \mathcal{A} = C^{\infty}(M) \otimes ( \mathbb{H}  \oplus  \mathbb{C}  \oplus  M_3(\mathbb{C} ) )
\end{equation}

The internal Hilbert space is: $\mathcal{H}
= \mathcal{H}_L \oplus \mathcal{H}_R \oplus \mathcal{H}_L^c \oplus \mathcal{H}_R^c$, where

\begin{displaymath}
   \mathcal{H}_L = ( \mathbb{C}^2 \otimes \mathbb{C}^N \otimes \mathbb{C}^3 )
   \oplus ( \mathbb{C}^2 \otimes \mathbb{C}^N )
\end{displaymath}

\begin{displaymath}
 \mathcal{H}_R = ( (\mathbb{C} \oplus \mathbb{C} ) \otimes \mathbb{C}^N \otimes \mathbb{C}^3 )
   \oplus ( \mathbb{C} \otimes \mathbb{C}^N )
\end{displaymath}

and whose basis is labelled by the elementary fermions and their antiparticles. The symbol $c$ is
used to indicate the section represented by the antiparticles. In either case of $\mathcal{H}_L$
and $\mathcal{H}_R$, the first direct summand is the quarks and the second, the leptons. $N$ stands
for the number of generations.  For example, the left-handed up and down quarks form an isospin
doublet and their right-handed counterparts are singlets and there are three colours for quarks and
none for leptons. The charges on the particles are identified by the faithful representation of the algebra on the Hilbert space.

Since the fermions can be left or right-handed, $\mathcal{H}$ is $\mathbb{Z}/2$ graded and the
grading operator is given by the chirality operator \mbox{$\chi$ = diag$(1,-1,1,-1)$}. If $\chi$
anticommutes with $\mathcal{D}$ then the triple is called even, otherwise it is called odd.
$\mathcal{H}$ may be written as a direct sum of two orthogonal parts represented by particles and
antiparticles. If it possesses the additional grading operator $\gamma$, which commutes with each
of $\mathcal{D}$, $\chi$ and $\rho(a)$ and whose eigenvalues are $+1$  for particles and $-1$ for
antiparticles and is called $S^o$-real. An $S^o$-real triple excludes Majorana spinors. The
standard model excludes Majorana particles and is formulated as a real, $S^o$-real, spectral
triple.

For a real spectral triple, there is an antilinear isometry $J$. The action of $J$ on $\mathcal{H}$ as given by the composition of charge conjugation and complex
conjugation:

\begin{displaymath}
   J \binom{ \psi_1}{\bar{\psi_2}}=\binom{\psi_2}{\bar{\psi_1}},
        \quad (\psi_1,\bar{\psi_2}) \in  \mathcal{H} \oplus \mathcal{H}^c
\end{displaymath}

where the bar indicates complex conjugation.

$J$ obeys the following relations for $KO$-dimension 6 and 0: $J^2=1$, $JD=DJ$. For $KO$-dimension 0, $J$ commutes with chirality $\chi$ and in the case of $KO$-dimension 6, $J$ anticommutes with $\chi$. The last two data correspond to the fact that in the Euclidean signature the chirality is unaffected by charge conjugation, while in the Lorentzian signature it is.

\section{Field content} \label{field content}

\subsection{Introduction}

Before we can start the calculations indicated in the introduction, we first have to establish exactly what is the field content of the theory both on and off-shell in each signature, and underline that the new fields do not conflict with experiment: we check that colour symmetry breaking does not occur, and that the additional fields do not add any new symmetries to the action. In this part we also show that new fields can provide Majorana terms that are non-zero in the physical vacuum. We will treat the Lorentzian and the Euclidean signatures in turn. For the latter case, we discuss the Wick rotation. The calculations in the two articles referred to in this introductory subsection were carried out in the Euclidean signature.

Paschke, Scheck and Sitarz showed that the new fields they called ``leptoquarks'' arise as those elements of the Dirac operator allowed by the first order condition when the $S^o$-reality condition is omitted. The effect of $S^o$-reality is to force the blocks labelled $Z$ below to 0 in order to disallow Majorana terms as in the standard model. The leptoquarks are elements of $Z$. 

\begin{equation}
  \left(  \begin{array}{cc}
Y        &     Z\\
\bar{Z}  &   \bar{Y}
 \end{array}  \right)
\end{equation}

where the  bar indicates complex conjugation, and the basis is the particles and antiparticle sections of the finite dimensional Hilbert space. \footnote{As the matrix given here is square, the basis really includes $\nu_R$, although a row and column of zeroes can always be written in for standard model considerations.} 

These new fields comprise 6 complex scalars having both the colour and weak isospin charges. They are due to the following connections between the quarks and leptons in the calculation of the first order condition as a constraint on $D_F$:                                                                                   This calculation has already been demonstrated fully in the papers we refer to, and to avoid needless repetition we will not give it here. The generalised $D_F$ without the $S^o$-reality condition is presented in the case of each signature below.

The generalisation given in \cite{smv} comprised of omitting $S^o$-reality, and also including the $\nu_R$ fermions. \footnote{To avoid violating Poincar\'e duality the chirality right $\tau$ neutrino was left out.} Two more new fields appeared for each fermionic generation, specifically a weak isospin doublet. As clarified below, (\ref{Euclidean}) this extension resulted in Majorana-type terms for the neutrino.

\subsection{Lorentzian}

``A Lorentzian version of the non-commutative geometry of the standard model of particle physics'' was worked out by Barrett \cite{lng} and similarly by Connes \cite{snm} in 2006. In this model the generalisation done in \cite{smv} by adding the $\nu_R$ and omitting $S^o$-reality was carried through to the Lorentzian signature. Since leptoquarks appear also in this model, they had to be ruled out to avoid colour symmetry breaking: in \cite{gsn} a constraint was applied to the Dirac operator to delete their presence from the action.

From the following axioms \cite{sap} in 0 dimensions: $D_F J = J D_F$, $D_F \chi = -\chi D_F$, $D_F = D_F^{\ast}$, $[[D_F, a],b^o]]$, $[[D_F, b^o],a]]$, one arrives at the following complex matrix presentation for $D_F$ with Hilbert space basis ($\Psi = \psi_L, \psi_R, \psi_L^c, \psi_R^c$), where each $\psi$ represents the set of standard model fermions with the addition of course of $\nu_R$:

\begin{equation}        \label{D}
D_F =
\left(   \begin{array}{cccc}
0         &         M^{\ast}    &   0        &   0\\
M         &          0          &   0        &   H\\
0         &          0          &   0        & M^T\\
0         &         H^{\ast}    &   \bar{M}  &  0
\end{array}  \right)
\end{equation}

where $M = M_Q \otimes 1_3 \oplus M_L$, that is, the Dirac mass matrix for quarks $Q$ and leptons $L$ and where $H$ is symmetric. In symbols, $H=H^t$ and $\bar{H}=H^{\ast}$ where $t$ and $\ast$ denote transposition and hermitian conjugation respectively. Considering the basis, we  note that the blocks labelled $M$ and $M^{\ast}$ are maps between left and right and they give Dirac masses by the Higgs mechanism, whereas $H$ is a map from right to right, which is indicative of Majorana mass terms. (Note that in the Lorentzian signature, $J$ anticommutes with $\chi$, and so in this case, $\chi=\textrm{diag}(1,-1,-1,1)$).

After applying all the commutation rules above involving $D_F$ (plus the new constraint), it is seen that for each family of fermions, there is only one non-zero entry which appears in the diagonal place in the final row and column (in the convention of \cite{smv}), which we label `$p$' and it arises due to the connection between chirality right neutrinos. When $D_F$ is substituted into the Yukwawa action terms: $(\Psi, D_F \Psi)$, the field $p$ gives the Majorana mass term: $\bar{\nu_R}~ p~ \nu_R$ and not surprisingly, from $\bar{p}$ arises the hermitian conjugate term.  If $\nu_R=0$ as in the standard model, $p$ is 0. Since $p$ only gives a mass to the $\nu_R$, the $\nu_L$ remains massless. 

Next we check that the field $p$ is non-zero in the physical vacuum. The reason for asking this question is that the solutions to the internal space Einstein's equations in \cite{smv} identifies zero and non-zero masses, and if we find that $p$ is zero by this method, it would not provide a non-zero mass in the Einstein equations calculated by the method in \cite{Schucker} either. So to check that  $\bar{\nu_R}~ p~ \nu_R$ be a valid Majorana mass term, we make this verification. To do this it is enough to consider the first generation alone.

As before, to calculate internal space Einstein's equations we vary the finite spectral action: \footnote{This is obtained by taking one point only in space-time, and setting the gauge fields to zero. With $D_F$ the standard model fermion mass matrix, it is equivalent to the usual Higgs potential.}

\begin{equation} \label{fsa}
S = \textrm{Tr}(D_F^4 - 2D_F^2)
\end{equation}

with respect to the degrees of freedom in $D_F$. 

Provided there are only quadratic terms or higher order, it doesn't matter whether we identify the zero elements of $H$ before or after performing the differentiation. The action $S$ is contains no lower order terms than quadratic terms in both $H$ and $M$.  

\begin{equation}    \label{H action}
 S = \textrm{Tr}[ 2(MM^{\ast})^2 + (H \bar{H})^2 + 4MM^{\ast} H \bar{H} -2(H\bar{H} + 2MM^{\ast})]
\end{equation}

This was obtained by inputting \ref{D} into \ref{fsa} and simplifying by using cyclicity of trace and the trace of a matrix is the same as the trace of its transpose and finally by writing the action as one part plus its hermitian conjugate and taking only the first part.

Moreover the equations of motion can be calculated just by differentiating \ref{fsa} with respect to $D_F$ to get:

\begin{equation}
 D_F^3 = D_F
\end{equation}

which tells us that:

\begin{eqnarray}
 D_F D_F^{\ast} D_F = D_F \\
 D_F^{\ast} D_F D_F^{\ast} = D_F^{\ast}
\end{eqnarray}

in other words the vacuum $D_F$ is a partial isometry. It has eigenvalues in the set {-1, 0 and +1} and so the diagonal form of $D_F^2$ is a diagonal matrix each of whose elements are 1 and 0. Therefore we write out the matrix equation $D_F^2$:

\begin{equation}        \label{D^2}
D_F^2 =
\left(   \begin{array}{cccc}
M^{\ast}M         &         0               &   0               &  M^{\ast} H \\
0                 &    MM^{\ast}+H\bar{H}   &   H\bar{M}        &    0\\
0                 &          M^t\bar{H}     &   M^t\bar{M}      &     0\\
\bar{H}M          &         0               &    0          &    \bar{H}H + \bar{M}M^t
\end{array}  \right) 
\end{equation}

It is clear that the field $p$ does not enter the equations involving the quark Dirac mass matrix and from the corresponding calculation done in \cite{tsm} it is immediate that the vacuum solution of the latter is a partial isometry. We can now work with only the first generation lepton basis and so $M$ and $H$ can be taken to be 2 by 2 complex matrices.

\begin{equation*}
M_L = \left(   \begin{array}{cc}
           q & r\\
	s & t
          \end{array}
\right), \qquad
M_Q = \left(   \begin{array}{cc}
           a & b\\
	c & d
          \end{array}
\right), \qquad
H = \left(   \begin{array}{cc}
           0 & 0\\
	0 & p
          \end{array}
\right)
\end{equation*}

Performing a gauge transformation so that each mass matrix $M$ and $H$ is diagonal, we find the vacuum solutions: $H$ and $M$ are orthogonal complement diagonal projections. Therefore the following two cases are solutions:

\begin{equation*}
 M=\left(  \begin{array}{cc}
1 & 0 \\
0 & 0
\end{array} \right), \qquad
H=\left(  \begin{array}{cc}
0 & 0 \\
0 & 1
\end{array} \right)t \bar{t}
\end{equation*}

and

\begin{equation*}
 M=\left(  \begin{array}{cc}
1 & 0 \\
0 & 1
\end{array} \right), \qquad
H=\left(  \begin{array}{cc}
0 & 0 \\
0 & 0
\end{array} \right)
\end{equation*}

The first solution given above tells us that the field $p$ (bottom right hand entry of $H$) has a non-zero vacuum solution and $\nu_R$ has a Majorana mass and no Dirac mass. In the other solution given above, $\nu_R$ has a Dirac mass and no Majorana mass. Seeing as $H$ is a field  independent altogether to $M$, the neutrino mass is unrelated to the Dirac masses of the other fermions. 

The pertinent action terms (\ref{action terms} below), followed by the resulting equation of motion for the field $t$ (\ref{t equation}) are:

\begin{equation} \label{action terms}
 (~t \bar{t}~)^2 + 2 ~t \bar{t}~(-1 +  r \bar{r} +  s \bar{s} + p\bar{p}) + 2\bar{r}\bar{s} q t
\end{equation}

\begin{equation}   \label{t equation}
 \bar{t} ( \vt t \vv + \vt r \vv + \vt s \vv + \vt p \vv - 1 ) + \bar{r}\bar{s}p = 0
\end{equation}

\underline{Conclusion:} the field content of the theory is that of the original spectral action principle plus the singlet complex scalar $p$. There are no contradictions to experiment as there are no leptoquarks to cause colour symmetry breaking and $p$ generates a Majorana mass in a similar way to the see-saw mechanism. Also, the introduction of these new scalar fields does not change the symmetry group of the action as the terms in the singlet complex scalar field $p$ are $U(1)$ invariant.

\subsection{Euclidean}  \label{Euclidean}

All of the calculations above have already been applied in the case of the Euclidean signature, so there is little original work to be done for this subsection. Nevertheless there are some important clarifications to be made before moving onto the following section.

In \cite{smv} the $\nu_R$ was added to the fermions and the $S^o$-reality condition was omitted. As a result, the axioms constraining the finite triple Dirac operator $D_F$ led to the appearance of the leptoquarks that had already made their debut in \cite{leptoquarks} plus the two new fields $j$ and $l$ that arose from the first order calculation oweing to the connection between $e_L^c$ and $\nu_R$, and between $\nu_L^c$ and $\nu_R$ respectively. They form a weak isospin doublet. Again we will not demonstrate here the first order condition derivation of the new fields because it already well documented. We only state the result of the calculation:

\begin{equation}
D_F = 
\left(   \begin{array}{cccc}
0         &       M^{\ast}     &     0      &    G   \\
M         &       0            &     G^T    &    0   \\
0         &   \bar{G}          &     0      &    M^T \\
G^{\ast}  &       0            &  \bar{M}   &    0
\end{array}   \right)
\end{equation}

with $M = M_Q \otimes 1_3 \oplus M_L$ as before. Notice that $G$ and $G^t$ are maps from $\psi_L$ to $\psi_R^c$ and $\psi_R$ to $\psi_L^c$ respectively indicating Majorana type terms. Unlike $H$ in the previous discussion, $G$ is not a symmetric matrix:

\begin{equation}
 G =
   \left( \begin{array}{cccccccc}
0 & 0 & 0 & 0 & 0 & 0 & 0 & 0\\
0 & 0 & 0 & 0 & 0 & 0 & 0 & 0\\
0 & 0 & 0 & 0 & 0 & 0 & 0 & 0\\
0 & 0 & 0 & 0 & 0 & 0 & 0 & 0\\
0 & 0 & 0 & 0 & 0 & 0 & 0 & 0\\
0 & 0 & 0 & 0 & 0 & 0 & 0 & 0\\
x & u & g & 0 & 0 & 0 & 0 & j\\
y & v & h & 0 & 0 & 0 & 0 & l
\end{array} \right)
  \end{equation}

where the basis is the set of chirality right antifermions in one family: six quarks and two leptons. Considering the basis and the first order condition result, we see that the leptoquarks ($g$, $u$, $x$, $h$, $v$, $y$) should appear in the bottom left-hand corner of the matrix and that the doublet $(j,l)$ appears in the bottom right.

By cyclicity of trace and freedom to take the transpose, the finite spectral action \ref{fsa} is equal to:

\begin{equation}            \label{M and G action}
S = \mathrm{Tr}  \big(  -2( G^{\ast} G  +  M M^{\ast}  )  +  (M M^{\ast})^2  +
(G^{\ast} G)^2 \\
+ 2( M G G^{\ast} M^{\ast}  +  M M^{\ast} G^T \bar{G}  +  M G
\bar{M} \bar{G})   \big)
\end{equation}

We see then that the leptoquarks will appear in the action unlike in the case of the Lorentzian signature, but they turned out to be zero in the physical vacuum \cite{smv}. In the symmetric phase vacuum where $M=0$, the leptoquarks are non-zero. 

In \cite{tsm} it was demonstrated that if the leptoquarks are non-dynamical, then $M$ in the vacuum has many features compatible with the standard model. In particular, it has a set of non-degenerate eigenvalues and $MM^{\ast}$ is diagonal while $M^{\ast}M$ is not diagonal.

(Since they are constant, the leptoquarks do not bring into effect a spontaneous breaking of colour symmetry). At first it seems unnatural that the leptoquarks be non-propagating, but these results seem unlikely to be just coincidental; and in the final section of this article, we suggest that the embedded strings produce a non-propagating torsion on the internal space and that this torsion can be associated with their stabilising fields, which are the leptoquarks.

There are solutions with $M \neq 0 $ and either $j$ or $l$ non-zero so they have the status of dynamical fields and they may generate the Majorana type terms $e_L^c \bar{j} e_R$, $e_R^c j^{\ast} e_L$ and $\nu_L^c \bar{l} \nu_R$, $\nu_R^c \bar{l} \nu_L$ (plus hermitian conjugates). This shows us that the value of the Majorana mass of the $\nu_L$ must be equal to that of the $\nu_R$. We also find that this mass term for the $\nu_L$ would not occur were it not for the $\nu_R$ since $l$ is only non-zero because $\nu_R$ is there. These two results seem to be unphysical, but we are working in the Euclidean signature, so we have to remember to Wick rotate to the physical spinors after the calculations. Since one of the aims of this section is to check for physical relevance, we explain that here.

Unlike the Dirac mass which is a map from left to right, the Majorana mass is a map from left to left or right to right:

\begin{equation}   \label{Mink}
m_M ( \Psi^{\ast}_L \Psi_L ) = m_M( \psi^{\ast}_L \psi^c_R + \psi_R^{\ast ~ c} \psi_L )
\end{equation}

where the second term is clearly the hermitian conjugate of the first term.
It has a similar right-handed counterpart. It is clear that the right-handed neutrino
$\Psi_R$ does not need to exist for the left-handed neutrino to get a Majorana mass.

In contrast, a Euclidean Weyl Majorana spinor is given by: $\Psi_L = \psi_L + \psi_L^c$. And
its right-handed counterpart is: $\Psi_R = \psi_R + \psi_R^c$. In this case of the Euclidean
signature, $J$ commutes with
$\chi=\mathrm{diag}(1,-1,1,-1)$ which means that changing a particle into an antiparticle and vice versa does not affect the chirality, and so if we apply the charge conjugation operator to $\Psi_R$ or to $\Psi_L$ we find that the Majorana condition is satisfied. The left-handed spinor of the dual basis is: $\Psi^{\ast}_L = \psi_L^{\ast} + \psi^{\ast ~ c}_L$. The Euclidean Majorana neutrino terms given above are equal to the product: $m_M \Psi^{\ast}_L \Psi_R$ plus hermitian conjugate: $m_M \Psi^{\ast}_R \Psi_L$ if $\Psi_L$ and $\Psi_R$ are left and right Euclidean neutrino fields. This shows that both chiralities must be present for a Euclidean Majorana term. Now let us compare the Minkowski and Euclidean terms.

The Euclidean Majorana  terms (which are generated by the map $l$:

\begin{eqnarray}
m_M \Psi^{\ast}_L \Psi_R = m_M ( \psi^{\ast}_L \psi^c_R + \psi^{c ~ \ast}_L\psi_R )\\
m_M \Psi^{\ast}_R \Psi_L = m_M ( \psi^{\ast}_R \psi^c_L + \psi^{c ~ \ast}_R\psi_L )
\end{eqnarray}

if we take the first term from the first row and the second term from the second row and
perform a Wick rotation then we get \ref{Mink}. The remaining terms give the right-handed
counterpart.

The results from the first order condition calculation show that $\nu_L$ may not acquire a
Majorana mass if the right-handed neutrino is not present because the existence of $l \neq 0$
requires that the representation for $\nu_R$ connects with the representation of the algebra on the antiparticle leptons. A physical reason for this could be that the
Minkowskian Majorana (Weyl) spinor as stated above comprises a projection onto both spin states and if $\nu_R$ is missing in this Euclidean framework, all the neutrinos would be left-handed: $\nu_L$ and its antiparticle $\nu_L^c$ and so the Euclidean neutrino would not Wick rotate into a physical Majorana spinor. 

To make a physical interpretation of the internal space Euclidean Majorana terms for $\nu_L$ and $\nu_R$ found above, we would make the identification back up
to the Euclidean space-time spinors (in the same vein as the identification due to fermion doubling) and then perform a Wick rotation of the Euclidean spinors to the physical Minkowski spinors. (In \cite{lng} the operator $J$ was redefined for the Lorentzian signature and the entire calculation of general $D_F$ was done from the beginning in the Lorentzian instead of doing the Wick rotation at the end of all the calculations.)

\underline{Conclusion:}  the field content is the same as that of the original spectral action principle plus the leptoquarks and plus the weak isospin doublet $(j,l)$. Hence, the action terms in this doublet are $SU(2)$ invariant. The leptoquarks do not cause colour symmetry breaking because they are not dynamical. Since $j$ and $l$ have no colour-charge, they cannot cause colour symmetry breaking. It is assumed that the leptoquark Majorana-type terms have very small couplings.

\section{Quantum and finite temperature corrections in $\phi^4$ theory}

In this section we follow the Coleman-Weinberg mechanism \cite{cw} for the
calculation of quantum corrections in $\phi^4$ theory and give details of the Dolan and Jackiw finite temperature
corrections \cite{dj} in the same theory. The aim is to present a pedagogical explanation of these useful
procedures for their own sake, while preparing for the subsequent section in which we study the
Electroweak phase transition in the theory that has become referred to as noncommutative geometry
with neutrino mixing.

\subsection{Phase transitions}

Here we provide a general introduction to phase transitions and leave a more detailed decription of
phase transitions in physics until the opening of the next section.

 A continuous
cosmological phase transition occurs when the universe undergoes an adiabatic expansion so that the
temperature drops below a certain critical temperature $T_c$ while a scalar field acts as the order parameter, taking on a
non-zero vacuum expectation value randomly chosen from the range of possibles that is the vacuum
manifold. During a phase transition, topological and embedded defects can develop. As solitonic
solutions to equations of motion in the cosmological context, these defects analogous to those
found in laboratory experiments also occur in cosmological theories. These are discussed in the
final section. Phase transitions can occur continuously or discontinuously as explained below.

A general form for the potential $V$ of a real scalar field $\phi$ is:

\begin{displaymath}
V = \alpha \phi^2 + \beta \phi^3 + \gamma \phi^4
\end{displaymath}

where the constants $\alpha$ to $\gamma$ depend on the temperature, pressure and affect the vacuum
expectation value of the field. The shape of a $V-\phi$ plot is a parabola that narrows as $\phi$
increases. In the symmetric phase the expectation value of $\phi$ is zero whereas if $\alpha$
becomes negative, (this is the mass term) there is a second solution to the equation of motion and
$\phi$ gets a non-zero expectation value corresponding to its value in the new vacuum, which is the
new phase with reduced symmetry. If $\beta$ is non-zero then the phase transition occurs suddenly
and the transition is called first order, for reasons clarified in the next section. If $\beta=0$,
then the change occurs gradually and  this is called a second order phase transition. However, for a
certain range of physical parameters, a theory with no cubic or linear terms can still encounter a
first order phase transition.

At small $\phi$, the quadratic term is more important than the quartic term, whereas for large
$\phi$,  the significance is reversed. Due to a symmetry breaking the sign of $\alpha$ changes and
in the case of a gradual phase transition, the shape of the  plot at small $\phi$ changes into an
upside-down parabola while a new minimum develops at a non-zero value of $\phi$. In the case of a
sudden transition, the old minimum remains and is separated from the new minimum by a potential
barrier. The old minimum is called the `false vacuum' and is classically stable due to the barrier,
but if the new minimum is at a lower potential than the old, then it is energetically favourable for
the field to move into the new phase and it can quantum tunnel through the barrier to the `true
vacuum'. The greater this potential difference, the greater the probability that the phase
transition happens.

A first order transition proceeds by nucleation of bubbles of new phase matter inside that of the
old  phase. The bubble walls move outwards and increasing volumes of old phase are swallowed up and
converted into true vacuum. Colliding bubbles coalesce until all of space is filled with true
vacuum, and the process is done.

So, in zero temperature classical field theory, phase transitions do not take place because there
is no spontaneous symmetry breaking mechanism to drive them. The mass term does not change sign so
the vacuum state of the field remains zero-valued. In practice however, phase transitions as
outlined above, do occur due to spontaneous symmetry breaking. This means that the classical
picture is not complete and it is necessary to perform corrections to the potential by considering
the zero temperature quantum effects. Physically, quantum fluctuations of the vacuum may alter the
form of the potential so that the field can develop a non-zero vacuum expectation value. As a
result the vacuum loses a symmetry and the matter obeys a smaller set of conservation laws. At
finite temperature, ($T \neq 0$) phase transitions can be brought about by temperature
fluctuations. The two cases will be considered here in turn.

\subsection{The Coleman-Weinberg Mechanism}

In this subsection we will outline the calculation of the one-loop correction at zero temperature
to the  self-interaction potential of a single, real scalar field $\phi$ and show that it leads to
symmetry breaking. The need for the modification is due to quantum fluctuations: $\phi \rightarrow
\phi_{cl}+\eta$.

Spontaneous symmetry breaking resulting from quantum fluctuations can induce a first order phase
transition if  it is implemented by quantum tunnelling - not a continuous process. The kinetic
(derivative) terms of the Lagrangian are unaffected by these corrections. The alteration is to the
\emph{interaction} terms and the result is called the effective potential,
$V_{eff}=V_{classical}(\phi)+O(\hbar)$ Here we will follow the work of Coleman and Weinberg who computed these corrections for a real scalar field $\phi^4$ theory. The process of
symmetry breaking studied here is called the Coleman-Weinberg mechanism.

The theory we will consider throughout this part is that of a real scalar field. The potential is:

\begin{displaymath}
V=\frac{1}{2} m \phi^2 + \frac{\lambda}{4!} \phi^4
\end{displaymath}

Whose value is the free energy density with the expectation value $<\phi>$ of the field. $<\phi>$ is
the `order  parameter' which will be defined in the next chapter. To calculate the quantum
fluctuations we perform the substitution $\phi \longrightarrow \phi_{cl} + \eta$ into the action.
The generator of Green functions is $Z=e^{\frac{i}{\hbar}W[j]}$ Where W[j] is the generator of
connected Green functions. Completing the square simplifies the subsequent calculation. The result
is:

\begin{displaymath}
\frac{i}{\hbar}W[j]=\frac{i}{\hbar} \left( S[\phi_{cl}] + \int j \phi_{cl} \right) -\frac{1}{2}~ \textrm{Tr}~
\ln  (\Box + m^2 + \frac{\lambda}{2} \phi_{cl}^2)+O(\hbar)
\end{displaymath}

\begin{displaymath} =>
W[j]=S[\phi_0]+\int j \phi_0 + \frac{1}{2}i \hbar~ \textrm{Tr}~ \ln (\Box +m^2+ \frac
{\lambda}{2}\phi_{cl}^2) + O(\hbar^2) \end{displaymath}

Now we perform the Legendre transformation: (which is familiar from switching from Helmholtz to
Gibbs  free energy.) $S_{eff}=W[j]-\int j \phi_{cl}$ and we obtain:

\begin{equation}  \label{result}
S_{eff}=S_{cl}+\frac{i \hbar}{2} ~\textrm{Tr}~ \ln (\Box+m^2+\frac{\lambda \phi^2}{2})+O(\hbar^2)
\end{equation}

Since $\phi_{cl}$ is a constant, the derivative terms vanish and as the action is given  by $S= \int
L d^4 x$, the effective potential $V_{eff}=-\frac{1}{VT} S_{eff}$ where $VT$ is the volume of
space-time. The final term is of course negligibly small and we will ignore it from now on. The
infinite series of Feynman diagrams are obtained term by term from the expansion of the logarithm.

It is impossible to compute the full effective potential because that would require a  summation of
an infinite number of Feynman diagrams. Instead we can either take a low order of perturbation
theory and compute as many of those diagrams as possible or we can select one or two typical
diagrams from each of many orders.  To lowest order, (called the tree approximation) the Feynman
diagram is a cross: it has four legs and a vertex. It indicates the original term
$\frac{\lambda}{4!}\phi^4$. The four legs represent the four powers of $\phi$ and the vertex
represents the coupling constant. To first order, (one loop) there is an infinite sequence of
polygon graphs begun below. Their sum is called the one-loop radiation correction.

The number of external legs represents the number of $\phi$ powers which shows the number  of
particles taking place in the interaction. For each vertex we write $\lambda$ and integrate over
the four-momentum. We include the $\frac{1}{2}$ to account for bose statistics; this is to prevent
us from overcounting equivalent diagrams. Each internal line carries a massless propagator
$\frac{i}{k^2-i\epsilon}$. A factor of $\frac{1}{2n}$ is also to prevent overcounting: reflection
or rotation of the polygons leads to the generation of equivalent diagrams, not new contractions in
the Wick expansion.

The resulting interaction terms of the effective potential are as follows. The second term is  the
coupling constant counter term with coefficient $C$. The ubiquitous $\hbar$ of quantum theory also
appears with the i as a prefactor but is not shown here because we are working in natural units
where $\hbar=c$=1.

\begin{equation} \label{expansion}
V_{eff}=\frac{\lambda}{4!}\phi^4-\frac{C}{4!} \phi^4+i\int\frac{d^4k}{(2\pi)^4} \sum_{n=1}^\infty
\frac{1}{2n} \bigg( \frac{\frac{1}{2}\lambda \phi^2}{k^2+i\epsilon}\bigg)^n \end{equation}

The final term in \ref{expansion} (we will label as `$V_1$') looks like the formula for the expansion of a
natural  logarithm ln(1+x) for small x. Transforming to Euclidean space, $i k_0 \rightarrow k_4$,
we rewrite it as:

\begin{displaymath}
V_1 = \frac{1}{2}\int\frac{d^4k}{(2\pi)^4} \ln(1+\frac{\lambda\phi^2}{2k^2})
\end{displaymath}

By dimensional analysis (or power counting) we can see that it is divergent. Coleman and Weinberg
performed a cut-off regularisation at $k^2=\Lambda^2$ which gave:

\begin{displaymath}
V_1=\frac{\lambda \Lambda^2}{64 \pi^2}\phi^2+ \frac{\lambda^2 \phi^4}{256\pi^2}~ [ ~\ln \bigg( \frac{\lambda
\phi^2}{2\Lambda^2} \bigg) -\frac{1}{2}~] \end{displaymath}

It then remained to renormalise the mass. (The resulting expression depends on the choice of
renormalisation  conditions. The choice does not alter the physical theory.) When this was
performed, it became necessary to shift the fields which in turn prompted a redefinition of the
coupling constants. This led to the determination of $\lambda$ in terms of the coupling strength e.
The final result was:

\begin{equation}
V_1 =\frac{3 e^{4}}{64 \pi^2} \phi^4 ~[ ~\ln  \frac{\phi^2}{\sigma^2}  -\frac{1}{2}~]
\end{equation}
where $\sigma$ is the renormalisation scale.

With the inclusion of the radiative correction order $\hbar$ term $V_1$ into the potential it
becomes  $V_{eff}=V_{\textrm{classical}}(\phi)+O(\hbar)$ This quantum correction drives a breaking
of the symmetry of the Lagrangian. Without this term the the solution to the equation
$\frac{dV}{d\phi}=0$ is $\vert \phi \vert=0$ but the solution to $\frac{dV_{eff}}{d\phi}=0$ gives a
non-zero expectation value of the field.

Although the steps outlined above were carried out in the case of a single, real scalar field,
the  methodology can be applied to other theories. For a non-Abelian gauge theory with scalar
fields, gauge fields and fermion fields, the Coleman-Weinberg correction with the `zero-momentum'
renormalisation conditions is as follows:

\begin{displaymath}
V_1(\phi)=\frac{1}{64 \pi^2}~[~\textrm{Tr}~(~\mu^4 \ln\frac{\mu^2}{\sigma^2} ~) + 3 \textrm{Tr}~(M^4 \ln \frac{M^2}{\sigma^2}~)
 - 4 \textrm{Tr}~(~m^4 \ln \frac{m^2}{\sigma^2}~)~] \end{displaymath}

where$ \mu$, $M$ and $m$ are the scalar, vector (gauge fields) and spinor (fermion) masses respectively.

\subsection{Finite temperature corrections}

At finite temperature, temperature fluctuations are clearly far more significant than
quantum fluctuations. Corrections to the potential must be made to account for the interactions
between particles in an ideal gas at finite temperature. Dolan and Jackiw calculated the finite temperature corrections to the potential.

With the aid of \cite{rivers} we will reproduce their derivation here in detail.

Firstly we recall the result of the first order summation of the Feynman diagrams from \ref{result}:

\begin{equation}   \label{4}
V_1=\frac{1}{2}\int\frac{d^4k}{(2 \pi)^4} \ln(k^2+U''(\phi))
\end{equation}

which is divergent. Note that $U''=\mathcal{M}^2$, that is, the coefficient in the quadratic term.
($\hbar$ is set to 1).

By switching to Euclidean space and integrating out the zeroth component of the momentum (this step
is shown below), $k_0$ we obtain:

\begin{equation}           \label{5}
V_1 = \int\frac{d^3k}{(2 \pi)^3}[\frac{1}{2}\omega+T\ln(1-\exp(-\beta\omega))]
\end{equation}

where $\beta$ is the inverse temperature.

In the zero temperature limit studied above, the second term clearly vanished ($T=0$) and the
divergent first term was analysed. The second term which is due to a mixture of quantum and
temperature fluctuations is not divergent. In this case of high temperatures, we omit the first
term which represents the zero point energy. At zero temperature, quantum fluctuations are
insignificant compared to variations in temperature so that changes in the first term are much
smaller than changes in the second, although some authors have considered both terms
simultaneously.

The steps to be taken to derive equation \ref{5} above from \ref{4} are as follows:

We impose the following boundary condition upon the Euclidean time coordinate:  `Time
period'=$\frac{1}{2 \pi T}$ where $T=$temperature and $\hbar$ is set to 1.

In this step, we have not replaced \emph{ physical time} with inverse temperature, and there is no
physical interpretation of the relation. After all, the Euclidean time coordinate is
\emph{imaginary}, having no physical significance. The quantities  to calculate in thermal quantum
field theory are thermal expectation values of observables, $<A>$. These are normally independent
of time and so the assignment of time dependence does not affect the physics. This provides a
freedom of choice that can be exploited by treating the density matrix as a time evolution
operator. This is explained below. \\The time-ordered thermal Green functions are defined as:

\begin{displaymath}
iG_t^{(N)}(t_1,...,t_N;T)=\frac{1}{Z} \textrm{Tr} \{ e^{-\beta H} T[\phi_1(\tau_1)... \phi_N(\tau_N)] \}
\end{displaymath}

The thermal expectation values differ from those calculated in zero-temperature quantum  field
theory as the time ordered product involves the density matrix, $e^{-\beta H}$ and the trace $(Tr)$
over all the possible states in the system. This brings the thermal average of statistical physics
into the quantum theory. The density matrix looks like the time evolution operator in Quantum
Mechanics. If we make the replacement:

\begin{displaymath}
e^{-\beta H} \rightarrow \exp \bigg(-\int_{\tau_i}^{\tau_i-i \beta} d \tau \hat{H} \bigg)
\end{displaymath}

$\beta$ defines the periodic boundary conditions on the imaginary time coordinate. Let $\tau_i=0$
then  $0 \leq \tau \leq \beta$. For fermions, the corresponding boundary conditions are
anti-periodic.

\begin{displaymath}
k_0=N\hbar\omega=\hbar \sum_n \frac{2 \pi}{\textrm{time period}}=2 \pi \sum_n 2 \pi T
\end{displaymath}

One of the factors of $2 \pi$ cancels with that in the invariant measure of $V_1$ when we  integrate
out the $k_0$:

\begin{displaymath} \int dk_0 \longrightarrow 2 \pi T \sum_{n=-\infty}^{n=+\infty}
\end{displaymath}

With this transformation the finite temperature correction becomes:

\begin{displaymath} V_1=\frac{1}{2}T \sum_n \int \frac{d^3k}{(2 \pi)^3}
\ln(k_i^2+\omega_{2n}^2 + \mathcal{M}^2)
\end{displaymath}

 where $\omega_{2n}^2=(2 \pi nT)^2$ and $k_i^2 + \mathcal{M}^2=E^2$

Let
\begin{displaymath}
{v(E)=\sum_n \ln(E^2+\omega_{2n}^2)}
\end{displaymath}
\begin{displaymath}
\frac{\partial v(E)}{\partial E}
=2 \sum_n \bigg(\frac{E}{E^2+ \omega_{2n}^2} \bigg)
\end{displaymath}

\begin{displaymath}
=2 \sum_n \bigg( \frac{E}{E^2+ (2 \pi n \beta^{-1})^2} \bigg)= \frac{2}{E} +  4
\sum_{n=1}^{\infty}\frac{E}{E^2+(2 \pi n \beta^{-1})^2} \end{displaymath}

\begin{displaymath}
=\frac{2}{E} + \frac{4 \beta}{2\pi} \sum
\frac{\frac{E\beta}{2\pi}}{\frac{E^2{\beta}^2}{{2\pi}^2}+n^2}   = \frac{2}{E}  +\frac{2\beta}{\pi}
\bigg( \frac{-\pi}{E\beta}+\frac{\pi}{2}\coth(\frac{E\beta}{2}) \bigg) \end{displaymath}

\begin{displaymath}
=\beta(\frac{e^x+e^{-x}}{e^x-e^{-x}})
=\frac{\beta e^{-x}(e^{2x}+1)}{e^{-x}(e^{2x}-1)}=\frac{\beta(e^{2x}+2-1)}{e^{2x}-1}
\end{displaymath}

\begin{displaymath}
=2\beta \bigg( \frac{1}{2}+ \frac{1}{e^{\beta E} -1} \bigg)
\end{displaymath}

~~~since $x=\frac{\beta E}{2}$

Let us restate this result more clearly:

\begin{displaymath}
\frac{\partial v(E)}{\partial E}=2\beta \bigg( \frac{1}{2}+\frac{1}{e^{\beta E}-1} \bigg)
\end{displaymath}

\begin{displaymath}
\int_{-\infty}^{+\infty} \frac{\partial v}{\partial E}dE =v(E)
=\int2 \beta \bigg( \frac{1}{2}+\frac{1}{e^{\beta E}-1} \bigg) dE
\end{displaymath}

\begin{displaymath}
=2\beta[\frac{1}{2}E+\beta^{-1}\ln(1-e^{-\beta E})]
\end{displaymath}

which is the result we were looking for. The last step can easily be checked by carrying it  out
backwards. That is to say differentiating the last expression gives the integrand in the
penultimate.

The final procedure is to write \ref{5} in a form in which we can easily read how these finite
temperature corrections can effect a symmetry breaking. If we assume that $\mathcal{M} \ll T$ we can
perform an expansion in powers of $\frac{\mathcal{M}}{T}$. These steps are set out below.

First we introduce the function $I(y)=\int_{0}^{\infty} dx~ x^2 \ln[1-e^{({x^2+y^2})^{\frac{1}{2}}}]$
where $y=\beta \mathcal{M}$ and so $V_1=\frac{I(\beta \mathcal{M})}{2 \pi^2\beta^4}$\\
$d^3k \longrightarrow dx$, so to ensure we do not make a dimensional error we include \mbox{the $x^2$.}\\
Taylor expanding:

\begin{displaymath}
I(y)=I(0)+yI'(0)+\frac{y^{2}I''(0)}{2!} +O(y^3)
\end{displaymath}

Evaluating the terms:
\begin{displaymath}
I(0)=I_0=\int_{0}^{\infty} dx~ x^2\ln(1-e^{-x})=-\sum_{n=1}^{\infty} n^{-1} \int dx~ x^2~ e^{-nx}=-\frac{\pi^4}{45}
\end{displaymath}

\begin{displaymath}
I'(0)=0 \qquad let~~ \frac{y^2I''(0)}{2} = y^2I_2
\end{displaymath}

\begin{displaymath}
\frac{dI(y)}{dy}=\int_{0}^{\infty} dx~ x^2[1-e^{-(x^2+y^2)^{\frac{1}{2}}}]^{-1} \frac{1}{2}
2y(x^2+y^2)^{-\frac{1}{2}}e^{-(x^2+y^2)^{\frac{1}{2}}} \end{displaymath}

\begin{displaymath}
\frac{d^2I(y}{dy^2}  \longrightarrow \int_{0}^{\infty} dx~ x^2[1-e^{-(x^2+y^2)^{\frac{1}{2}}}]^{-1}
(x^2+y^2)^{-\frac{1}{2}}e^{-(x^2+y^2)^{\frac{1}{2}}} \end{displaymath}

evaluating at $y=0 \ldots$
\begin{displaymath}
\int_{0}^{\infty} dx x^2 (1-e^{-x})^{-1} \frac{1}{x} e^{-x}=\int_{0}^{\infty} dx ~x~e^{-x}(1-e^{-x})^{-1}
\end{displaymath}

\begin{displaymath}
\frac{y^2I''(0)}{2}=y^2I_2
\\=\frac{1}{2}\int_{0}^{\infty} dx~ x~ e^{-x}(1-e^{-x})^{-1}=\frac{\pi^2}{12}
\end{displaymath}

we substitute these back into $V_1=\frac{I(\beta M)}{2 \pi^2\beta^4}$ to obtain the temperature
dependent contribution to the free energy:

\begin{equation}
V_1=-\frac{\pi^2}{90}T^4+\frac{\mathcal{M}^2}{24}T^2+O(T)
\end{equation}

with the replacement $y \longrightarrow \beta \mathcal{M}$

The $O(T)$ term is neglected in the high temperature limit.

The final effective potential becomes:

\begin{equation}  \label{7}
V_{eff}(\phi,T)=-\frac{m^2}{2}\phi^2+\frac{\lambda}{4}\phi^4-\frac{\pi^2}{90}T^4+\frac{\mathcal{M}^2}{24}T^2
\end{equation}

The physical interpretation of the last result now follows.
\\With $\mathcal{M}^2=\frac{d^2V}{d\phi^2}=3\lambda\phi^2-m^2$

the last equation \ref{7} becomes:
\begin{eqnarray}   \label{8}
V_1=-\frac{\pi^2}{90}T^4+\frac{T^2}{24}(3\lambda\phi^2-m^2) \nonumber=-\frac{\pi^2
T^4}{90}+\frac{\lambda  T^2}{8}\phi^2-\frac{m^2 T^2}{24} \nonumber
\\\\V_{eff}=-\frac{m^2\phi^2}{2}+\frac{\lambda \phi^4}{4}+\frac{\lambda T^2\phi^2}{8} \nonumber
\end{eqnarray} \\+terms not dependent upon $\phi$.

The field varies smoothly as the temperature changes. A temperature change from one side of the
critical  temperature to the other would induce a second order phase transition. For example, the
equation holds that if the temperature is increased up to the critical temperature, the field
decreases continuously to zero.

In general the $\phi^2$ term yields the mass squared. Inspection of this coefficient yields crucial
information about how a phase transition is implemented by a temperature change.

\begin{equation} \label{9}
\frac{\lambda T^2}{8}\phi^2-\frac{m^2}{2}\phi^2~=~ \frac{1}{2}[\frac{\lambda
T^2}{4}-m^2]\phi^2~=~\frac{1}{2}  \mathcal{M}^2 \phi^2 \end{equation}

Four points to consider about the above expression are:

\begin{itemize}
\item When $\mathcal{M}^2<0$ the field obtains a vacuum expectation value and the symmetry  is
broken;~~ $T<T_c$
\item When $\mathcal{M}^2>0$ the symmetry is maintained and $\phi=0;~~~~ T>T_c$
\item When the temperature $T$ increases from below $T_c$, $\mathcal{M}^2$ becomes =0 and then       it
becomes positive and symmetry is restored.
\item The critical temperature $T_c$ is defined by $\mathcal{M}^2=0$.
\end{itemize}

It is evident from the above equation that the approximation $T \gg \mathcal{M}$ we used in the
expansion carried out in the previous subsection is valid provided the coupling constant $\lambda$
is small.

\section{Analysis of the effective potential: determining the order of the phase transition in two theories}

In this section we will demonstrate how to determine the order of a phase transition  from the
effective potential in finite temperature field theory.
In the previous section, the finite temperature corrections were demonstrated only in
$\phi^4$-theory, but equation \ref{7} can be applied to any theory just by taking the second
differential of the potential in that theory: $U''$. It follows that we may write down the effective
potential for noncommutative geometry with neutrino mixing, and subsequently analyse it in
the Ehrenfest classification (see below) in order to predict whether the Electroweak phase
transition was to be first or second order.    We also assume
that after a Wick rotation back to the Lorentzian, we can still add on the correction terms in the
same way where the meaning of $T$ becomes physical temperature, relinquishing its connection with
imaginary time.

 The $\phi^4$ theory is used like a control  experiment as we already know that the answer is that
it be a second order phase transition because does not have a linear or cubic term. And these
derivations are carried out side by side for a clear and direct comparison. The
purpose of this section is to carry out a corresponding calculation in the case of noncommutative
geometry with neutrino mixing to ascertain whether the new fields in the theory change this. The
Electroweak phase transition is first order for a certain range of physical parameters, but we ask
whether the new fields ensure that it be first or second order for a greater range of parameters.
Of course, the new interactions are quadratic in the Higgs, so if the leptoquarks are non-dynamical,
we expect the phase transition to be second order. However, since as dynamical fields, they
interact with the Higgs in the vacuum, we should work through the calculation to find out whether
they have any effect on the order of the phase transition.

First we explain a little about the relevant physics.

The Ehrenfest classification (1933) of phase transitions states that if the nth derivative of the chemical potential  $\mu$ is the
lowest order discontinuous derivative then the transition is nth order. We will discuss
$\frac{dV}{dT}$ rather than $\frac{d\mu}{dT}$ since the Gibbs free energy $G$ and chemical
potential are related in the following simple way:

\begin{displaymath}
\mu=\frac{\partial G}{\partial N}=\frac{\partial (-kT \ln Z - \Omega)}{\partial N}~~~\textrm{implying}~~G=\int\mu dN
\end{displaymath}

For a \emph{first order} phase transition $\frac{dV}{dT}$ is discontinuous. The  heat capacity at
constant pressure $C_p$ is given by the derivative of enthalpy $H$ with respect to temperature.
This is also infinite since the enthalpy changes a finite amount for an infinitesimal change in the
temperature. For example, water heated to its boiling point remains at the same temperature as heat
continues to be applied to it since the energy is being taken up by the bond breaking rather than
to increase the kinetic energy of the particles.

In general, if there is a non-zero latent heat released during the phase change then the  transition
is \emph{first order}. When heating a liquid until it evaporates its pressure maintains a constant
value $p<p_c$ where $p_c$ is the pressure at the critical point. When bubbles of the new phase
form, the pressure jumps to its new value. $\Delta p$ has a simple and direct relationship with the
change in free energy. If the evaporation experiment is repeated but this time the pressure is held
at constant pressure $p=p_c$ the latent heat becomes smaller and in the limit, no heating is
required to convert the liquid into vapour. At the critical point, the latent heat of vaprization
is zero and as a consequence, the first derivative of the Helmholtz free energy with respect to
temperature is continuous at that point and so we have a second order phase transition.

In the case of a \emph{second order} phase transition the first derivative of the free energy  with
respect to temperature is continuous but the second order derivative is not. This time the heat
capacity at constant pressure changes discontinuously but is not infinite. An example of a second
order phase transition can be given as a semiconductor whose temperature increases beyond the
critical temperature and it becomes an insulator, or the recrystallisation of a metal by annealing.

Below we will use the Ehrenfest classification to analyse the effective potential (\ref{7}) in
finite temperature field theory and compare the forms of the potentials for first and second order
phase transitions.

\subsection{Analysis of the effective potentials}                \label{ftc}

The following cases are considered in parallel:

$(A)$: ordinary $\phi^4$ theory

$(B)$: noncommutative geometry with neutrino mixing in the Lorentzian signature.

$(C)$: noncommutative geometry with neutrino mixing derived from an irreducible (in  the sense of
\cite{smv}) spectral triple and with colour charge neglected, in the Euclidean signature.

$(C2)$: the same as $C$ but with constant leptoquarks.

$(D)$: the same as $C$ but in the theory with the full fermion basis to include their colour charge. 

$(D2)$: As $(D)$ but with constant leptoquarks.

The reason that the cases $(B)$, $(D)$ and $(D2)$ are each studied is explained in the
introduction and the notation is defined in \ref{field content}. Cases $(C)$ and $(C2)$ are also worked through in order to make the comparison to analyse the
effect if any, that the existence of colour charge may have on the derivation. Case $(A)$ is
included to act as a control experiment with which to compare the stated theories. For some formulae
and some steps, $(C2)$ and $(D2)$ are the same as $(C)$ and $(D)$ respectively, so we do not
always write them separately.

In writing down the potentials $V_{(A)}$ to  $V_{(D2)}$, \footnote{We need only to include the terms
in  $a$, $t$. For example, linear terms in $\bar{a}$ are excluded.} we will study only one row from
each set of matrix equations that are the equations of motion (integrated) derived from the finite spectral actions presented in \ref{field content} (for example \ref{t equation}) because for the purposes of the calculations to follow, it is enough to consider one row at a time. Not all the details as to the derivation of the potentials are given here because this information has already been presented and analysed in \cite{smv}. The solutions pertaining to the symmetric phase extremum of the
potential and that of the new phase, or physical vacuum are written down in the tables below. For
$(B)$ we will mainly focus on the solution where $p$ is zero in the physical vacuum. The field
labels, $t$, $a$ and so on have all been defined in \ref{field content}. The equations of motion in the
Lorentzian case were given earlier in \ref{field content}, whereas those for the Euclidean case were calculated in articles cited in that section. Let $I$ denote the symmetric old phase and $II$ the
new phase with less symmetry.

\bd
V_{(A)} = \frac{\lambda}{4} \phi^4 -\frac{m^2}{2} \phi^2
\ed

\bd
V_{(B)} = \vert t \vert^4 + 2\vert t \vert^2 (\vert r \vert^2 + \vert s \vert^2 + \vert p \vert^2 -1)
   +2\bar{r}\bar{s}qt  \ed

\bd
V_{(C)} = \vert a \vert^4 + 2\vert a \vert^2 ( -1 + \vert b \vert^2 + \vert c \vert^2 + \vert g
\vert^2 +  \vert h \vert^2 )   +2\bar{c} \bar{b}da  \ed

\bd
V_{(D)} = 3 \vert a \vert^4 + 6 \vert a \vert^2(\vert b \vert^2 + \vert c \vert^2 + \vert g \vert^2
+  \vert u \vert^2 + \vert x \vert^2 + \\\\ \vert h \vert^2 + \vert v \vert^2 + \vert y \vert^2 -3)
 + 6 \bar{c} \bar{b}da \ed

\vspace{1cm}
\begin{tabular}{|c|c|}
\hline
\\
     &   \underline{Symmetric phase  $I$}          \\\\
\hline
\\
$(A)$ & $\vert \phi \vert^2 = 0$  \\ \\
\hline
\\
$(B)$ & $\vert t \vert^2 = \vert s \vert^2 = 0$,~~ $\vert p \vert^2 = 1$  \\\\
\hline
\\
$(C)$ & $\vert a \vert^2 = \vert b \vert^2 = 0$, ~~$\vert g \vert^2 + \vert h \vert^2 = 1$ \\ \\
\hline
\\
$(C2)$ & $\vert a \vert^2 = \vert b \vert^2 = 0$, ~~ $g$,  $h$ constant,~~ $\bar{a}(\vert a \vert^2
+ \vert b \vert^2 + \vert g \vert^2 + \vert h \vert^2 - 1) = 0$ \\ \\  \hline  \\  $(D)$  & $\vert a
\vert^2 = \vert b \vert^2 = 0$,~~ $\vert g \vert^2 + \vert u \vert^2 + \vert  x \vert^2 + \vert h
\vert^2 + \vert v \vert^2 + \vert y \vert^2 = 3$ \\\\  \hline  \\  $(D2)$ & $\vert a \vert^2 =
\vert b \vert^2 = 0$,~~ $g$, $u$, $x$, $h$, $v$, $y$ constant \\\\  \hline  \end{tabular}

\vspace{1cm}

\begin{tabular}{|c|c|}
\hline
\\
     &   \underline{New phase  $II$}          \\\\
\hline
\\
$(A)$ & $\vert \phi \vert^2 = \frac{M^2}{\lambda}$ \\
\\
\hline
\\
$(B)$ & $\vert t \vert^2 + \vert s \vert^2 =1$,  ~~ $\vert p \vert^2 = 0$  \\
\\
\hline
\\
$(C)$ &  $\vert a \vert^2 + \vert b \vert^2 = 1$,~~ $g=h=0$ \\
\\
\hline
\\
$(C2)$ & $\vert a \vert^2 + \vert b \vert^2 + \vert g \vert^2 + \vert h \vert^2 = 1$,~~$a \neq
0$,~~$\vt a \vv \neq 1$\\ \\
\hline
\\
$(D)$  & $\vert a \vert^2 + \vert b \vert^2 = 1$,~~ $g=u=x=h=v=y=0$ \\
\\
\hline
\\
$(D2)$ & $\vert a \vert^2 + \vert b \vert^2 + \vert g \vert^2 + \vert u \vert^2 +
\vert x \vert^2 + \vert h \vert^2 + \vert v \vert^2 + \vert y \vert^2 = 3$~~ $a \neq
0$,~~$\vt a \vv \neq 1$\\ \\
\hline
\end{tabular} \vspace{1cm}

\subsubsection{Finite temperature corrections}  \label{V}

Now we test for what type of phase transition it may have been that happened as the Electroweak
epoch progressed into the electromagnetic epoch of today. In the way already explained, we write
down the finite temperature corrections for each case:

\begin{eqnarray*}
V_{eff~(A)} =  \frac{\lambda}{4} \phi^4 -\frac{m^2}{2} \phi^2 +  -\frac{\pi^2}{90} T^4 +
\frac{\mathcal{M}_A^2 T^2}{24}\\\\\\ V_{eff~(B)} = \vert t \vert^4 + 2\vert t \vert^2 ( -1 + \vert r
\vert^2 + \vert s \vert^2 + \vert p \vert^2  )  + \bar{r}\bar{s}qt -\frac{\pi^2}{90} T^4 +
\frac{\mathcal{M}_B^2 T^2}{24}\\\\\\ V_{eff~(C)} = \vert a \vert^4 + 2\vert a \vert^2 ( -1 + \vert b
\vert^2 + \vert c \vert^2 + \vert g \vert^2 + \vert h \vert^2 )  + 2\bar{c} \bar{b}da
-\frac{\pi^2}{90} T^4 + \frac{\mathcal{M}_C^2 T^2}{24}\\\\\\ V_{eff~(D)} = 3 \vert a \vert^4 + 6
\vert a \vert^2 ( \vert b \vert^2 + \vert c \vert^2) + 2\vert a \vert^2 (\vert g \vert^2 +\vert u
\vert^2 + \vert x \vert^2 + \vert h \vert^2 + \vert v \vert^2 + \vert y \vert^2 -3 ) \\\\ + 6
\bar{c} \bar{b}da -\frac{\pi^2}{90} T^4 + \frac{\mathcal{M}_D^2 T^2}{24} \end{eqnarray*}

where

\begin{displaymath}
\mathcal{M}_{(A)}^2=\frac{d^2V_{(A)}}{d\phi^2}
\end{displaymath}

and so on is given by:

\bd
\mathcal{M}_{(A)}^2 = 3 \lambda \phi^2 - m^2
\ed

\bd
\mathcal{M}_{(B)}^2 = 4\vert t \vert^2 + 2\vert r \vert^2 + 2\vert s \vert^2 + 2\vert p \vert^2 -2
\ed

\bd
\mathcal{M}_{(C)}^2 = 4 \vert a \vert^2 + 2\vert b \vert^2 + 2\vert c \vert^2 + 2\vert g  \vert^2 +
2\vert h \vert^2 -2  \ed

\bd
\mathcal{M}_{(D)}^2 = 6 (\vert a \vert^2 + \vert b \vert^2 + \vert c \vert^2) + 2(\vert g  \vert^2
+\vert u \vert^2 + \vert x \vert^2 + \vert h \vert^2 + \vert v \vert^2 + \vert y \vert^2 -3)  \ed

\subsubsection{Critical temperature}

Substituting for $\mathcal{M}$ in $V_{eff}$, collecting together the terms quadratic in  $\phi$ and
omitting terms in $T$ but not in $\phi$, $t$, or $a$ respectively to find the mass terms, while working in mass eigenstates, that is, $r=s=b=c=0$:

\begin{eqnarray*}
V_{eff~(A)} =  \frac{\lambda}{4} \phi^4 -\frac{1}{2} ( m^2 - \frac{\lambda T^2}{4}) \phi^2 \\\\\\
V_{eff~(B)} = \vert t \vert^4 + 2\vert t \vert^2 ( -1 + \vert p \vert^2) +  \frac{4T^2}{24} \\ \\\\ V_{eff~(C)} = \vert a \vert^4 + 2\vert a \vert^2( -1 +  \vert g \vert^2 + \vert h \vert^2) + \frac{4T^2}{24} \\\\\\ V_{eff~(D)} = 3\vert a \vert^4 + 2\vert a \vert^2( -3 + \vert g \vert^2 + \vert u \vert^2 + \vert x \vert^2 + \vert h \vert^2 + \vert v \vert^2 + \vert y \vert^2) + \frac{6T^2}{24}
\end{eqnarray*}

with masses:

\begin{eqnarray*}
-\frac{1}{2}M_{(A)}^2 = -\frac{1}{2} (m^2 - \frac{\lambda T^2}{4}) \\\\\\
-\frac{1}{2}M_{(B)}^2 = -\frac{1}{2}( 4 -   4\vt p \vv   - \frac{T^2}{3} )\\\\\\
-\frac{1}{2}M_{(C)}^2 = -\frac{1}{2}( 4 -   4\vt g \vv - 4\vt h \vv   -\frac{T^2}{3} )\\\\\\
-\frac{1}{2}M_{(D)}^2 = -\frac{1}{2}( 6 -   4\vt g \vv -4\vt u \vv -4\vt x \vv -4\vt h \vv - 4 \vt v \vv -4\vt y \vv   -\frac{T^2}{2} )
\end{eqnarray*}

As the temperature increases, one can see that the mass term goes positive,  where the only vacuum
solution for the field is 0. The critical event, where the mass squared is 0, is when the universe
is at the critical temperature, $T_c$. For constant $G$, the contribution to the critical temperature from the new fields is to decrease it, as $-\vt g \vv -\vt u \vv -\vt x \vv -\vt h \vv -\vt v \vv -\vt y \vv +3$  takes a values dependent on the square of the mass of the up particle. However, it is difficult at this stage to state anything definitive about actual numbers.

\subsubsection{The Ehrenfest classification}

Since we are making a comparison, we may omit any terms that occur
identically on both sides of the critical event.  Here, let us substitute with $\phi=0$ to write
down the potential (using \ref{V}  above) during the symmetric phase:

\bd
V_{I,~(A)} = V_{I,~(B)} = V_{I,~(C)} = V_{I,~(D)} = 0
\ed

\bd
V_{I,~(C2)} = -\frac{T^2}{12} \vert a_{II} \vert^2 =   \frac{T^2}{12} ( \vert g \vert^2 + \vert h
\vert^2 -1)
\ed

\bd
V_{I,~(D2)} = -\frac{T^2}{12} 3 \vert a_{II} \vert^2 =   \frac{T^2}{12} ( 2 (\vert g
\vert^2 + \vert u \vert^2 + \vert x \vert ^2 + \vert h \vert^2 + \vert v \vert^2 + \vert y \vert^2)
-6)
\ed

At $V_{II}$ the new phase, we substitute for the non-zero vacuum expectation value of the field:

The formula $\vert a \vert^2 + \vert b \vert^2$ is interpreted as  the Higgs doublet term $\vert
\phi \vert^2=\vt \phi_1 \vv + \vt \phi_2 \vv$ and as the true vacuum the Higgs chooses a direction at random we can take $\phi_2=0$,  so we choose $b=0$.
Similarly, let $s=0$.

\bd
V_{II,~(A)} =  -\frac{M^4}{4 \lambda}
\ed

\bd
V_{II,~(B)} =  V_{II,~(C)} =     \frac{T^2}{6};~~~~V_{II,~(D)} = \frac{T^2}{4}
\ed

\bd
V_{II,~(C2)}  = \frac{T^2}{6} \vert a \vert^2 +  V_{I,~(C2)} ~~~~= \frac{T^2}{12} \vert a \vert^2
\ed

\bd
V_{II,~(D2)} =  \vert a \vert^2 ( 3\vert a \vert^2 - 3 \vert a \vert^2) +
\frac{T^2}{12}(6\vert a \vert^2) +  V_{I,~(C2)} ~~~~=  \frac{T^2}{12}(3 \vert a \vert^2)
\ed

So, $V_{II,~(C2)}  =  -V_{I,~(C2)}$ and   $V_{II,~(D2)}  =  -V_{I,~(D2)}$.

Next we compare the  first differential of the two potentials with respect to the temperature
evaluated at the critical temperature:

\begin{displaymath}
\frac{dV_{I,~(A)}}{dT} = \frac{dV_{I,~(B)}}{dT} = \frac{dV_{I,~(C)}}{dT} =  \frac{dV_{I,~(D)}}{dT} =  0
\end{displaymath}

At $T=T_c$ the modulus of  $\frac{dV_{I,~(C2)}}{dT}$ and  $\frac{dV_{I,~(D2)}}{dT}$   is
equal to that of   $\frac{dV_{II,~(C2)}}{dT}$ and     $\frac{dV_{II,~(D2)}}{dT}$     respectively.

At the critical temperature:
\begin{eqnarray*}
\bigg( \frac{dV_{II,~(B)}}{dT} \bigg)_{T_c} = \bigg( \frac{dV_{II,~(C)}}{dT} \bigg)_{T_c} =  \frac{T}{6} \neq 0\\\\\\
 \frac{dV_{II,~(A)}}{dT} = \frac{dV_{II}}{dM^2} . \frac{dM^2}{dT} = \bigg( \frac{M^2T}{4} \bigg)_{T_c}= 0;~~~~~ \bigg( \frac{dV_{II,~(D)}}{dT} \bigg)_{T_c} = \frac{T}{2} \neq 0
\end{eqnarray*}

The results show that the potentials for $(B)$, $(C)$ and $(D)$ have a discontinuous first
derivative at $T=T_c$ and we conclude that the phase transition is first order.
Whereas for $(A)$, $(C2)$ and $(D2)$, the first derivatives are continuous at the critical event,
which means that the phase transition cannot be first order. For a second order phase transition,
the second order derivative  of the potential with respect to the temperature is discontinuous at
the critical temperature.  $\frac{d^2V_{I,~(C2)}}{dT^2}$ and  $\frac{d^2V_{I,~(D2)}}{dT^2}$  have
the same modulus but opposite signs to   $\frac{d^2V_{II,~(C2)}}{dT^2}$ and
$\frac{d^2V_{II,~(D2)}}{dT^2}$     respectively. We expect $V_{II}$ to decrease as the temperature
decreases and $V_{I}$ to increase as the temperature decreases, and since $T$ decreases as time
progresses, while the expectation value for the field increases, for a second order phase
transition we expect the second order differential with resect to temperature to behave in the same
way as the second order differential with respect to the field, which is, for the original minimum
to evolve into a maximum, and the new minimum to become progressively deeper until the phase
transition is complete. 

In the $\phi^4$-theory, the second differential of the potential with
respect to temperature is discontinuous at $T=T_c$:

\bd
\frac{d^2V_{II,~(A)}}{dT^2} = \frac{1}{4} \bigg( m^2 - \frac{3}{4} \lambda T^2 \bigg) \neq 0
\ed

\bd
\frac{d^2V_{I,~(A)}}{dT^2}=0
\ed

\subsubsection{Type of extremum in the symmetric phase}

In this part we will establish whether the turning point at $\phi=0$ is a maximum, a
minimum or a saddle point for each of the cases. To do this, we simply differentiate $V_{(A)}$ to
$V_{(D)}$ given in part \ref{ftc} twice to see how the gradient changes at that point.

\bd
  \frac{dV_{(A)}}{d\phi} = \frac{1}{2} \lambda \phi^3 - \frac{m^2}{2}\phi= 0\\
\ed

\bd
\frac{d^2V_{(A)}}{d\phi^2} \vert_{\phi=0}= - \frac{m^2}{2} < 0
\ed

\bd
\frac{d^2V_{(B)}}{dtd\bar{t}}\vert_{M_L=0} = 2(\vert p \vert^2 -1) = 0
\ed

\bd
\frac{d^2V_{(C)}}{d ad \bar{a}}\vert_{M_Q=0,~G=I} = 2(\vert g \vert^2 + \vert h \vert^2 - 1) = 0
\ed

\bd
\frac{d^2V_{(C2)}}{dad\bar{a}}\vert_{M_Q=0,~constant~G} = 2(\vert g \vert^2 + \vert h \vert^2 - 1) <0
\ed

The last line is not equal to zero because $a=0$ at $I$, and so the $\bar{a}$ doesn't cancel in the equation of motion: $\bar{a} ( \vt a \vv + \vt g \vv + \vt h \vv -1 )=0$ and since $g$ and $h$ do not have their own equations of motion, $\vert g \vert^2 + \vert h \vert^2 - 1 \neq 0$.

\bd
\frac{d^2V_{(D)}}{dad\bar{a}}\vert_{M_Q=0,~G=I} = 6(\vert g \vert^2 + \vert u \vert^2 +\vert x
\vert^2 + \vert h \vert^2 +\vert v \vert^2 + \vert y \vert^2 - 1) = 0 \ed

\bd
\frac{d^2V_{(D2)}}{dad\bar{a}}\vert_{M_Q=0,~constant~G} = 6(\vert g \vert^2 + \vert u \vert^2  +\vert
x \vert^2 + \vert h \vert^2 +\vert v \vert^2 + \vert y \vert^2 - 1) <0 
\ed

In the vacuum where $t=0$ and $\vt p \vv = 1$ in the new phase, and $p=0$ while $\vt t \vv = 1$ in the old phase, we get the same potentials as in $(B)$ except that the expressions for $V_I$ and $V_{II}$ are traded. And the two procedures given above can be used to show that the phase transition is first order. We can say something new about the critical temperature. Recalling that $p$ is the generating field of the $\nu_R$ Majorana mass, it is expected to take a large value, and hence its effect on the critical temperature is to reduce it significantly. This is further confirmation that the phase transition is \emph{weakly} first order.

These results show that the extremum at $I$ is a maximum in cases $(A)$, $(C2)$ and $(D2)$ whereas
it is a saddle point in the cases $(B)$, $(C)$ and $(D)$. The former confirms that the greater the
values of the constant leptoquarks, the greater the range of physical parameters for which the
phase transition is second order. The latter confirms that in the corresponding cases, the phase
transition is very weakly (in fact only just) first order. Comparing $(C)$ with $(D)$, we find that when colour charge is included, as there are even more new fields, the critical temperature is reduced even more, and that the maximum at $I$ is even steeper. It is $(B)$ and $(C)$ that are referred to in the abstract as these cases are already better understood and recognised than $(C2)$ and $(D2)$.

\subsubsection{Summary}

The calculations in this section led to the following predictions:

\vspace{1cm}

\begin{tabular}{|c|c|c|c|}
\hline \hline
\\
$\phi^4$-theory & Euclidean & Lorentzian & Euclidean,  \\
                 &           &               &       $G$ constant
\\
\hline \hline
\\
maximum   & saddle point  &  saddle point  &  maximum
\\
\hline
\\
2nd order   &  just 1st order   & just 1st order  & 2nd order \\
phase transition  & phase transition & phase transition  &  phase transition
\\
\hline \hline
\end{tabular}

\vspace{1cm}

It should be pointed out once again that the scope of these results is relative to the range of physical parameters employed. A more extensive study involving the space of physical parameters might lead to a more quantative estimate as to what extent the new fields can increase or decrease the likelihood that the Electroweak phase transition be first order. As mentioned, for a certain range of parameters it is already established in the literature that the Electroweak phase transition was probably first order. The aim of our study was to determine whether the effect of the new field interactions would be to make it more or less likely that this prediction be correct.

\section{Embedded defects and torsion}

In this section we show that the additional fields in the extension of the standard model in noncommutative geometry significantly improve the stability of the Electroweak $Z$ string. Using an analogy from condensed matter physics experiments, we suggest that the Electroweak $Z$ string solution in this extension is a source of torsion in the internal space dimension. Since the string is a static solution, this torsion is not expected to be dynamical, and this scenario is given as an explanation as to why the leptoquarks should not have equations of motion. The motivation for having constant leptoquarks was explained in the introduction. The reasons for introducing torsion in this way is that (a) the leptoquarks satisfy the formulation of torsion suggested below, which is constructed in analogy to that of Cartan, and (b) because of an argument given below in the context of internal space gravity, also called  the Higgs force. Finally, we discuss the relationship between the energetics of the string and internal space curvature and the metric on the vacuum manifold and find that there is a discrete set of these metrics, and hence string mode, for each fermion mass.

\subsection{Introduction to topological and embedded defects}

\subsubsection{What are topological and embedded defects?}

A soliton is a non-dissipating solitary wave. In other words, it ``keeps itself together'' like a pulse sent down a fibre optic cable whose energy remains localised due to the refractive index of the glass.\footnote{There is a sculpture in Lisbon that emits solitonic water waves} Solitons are particle-like solutions of the equations of motion. Topological defects are solitons with a non-trivial topology to which they owe their stability. They consist of localized trappings of old phase matter persisting on into the new phase, so they are a relic of symmetric phase matter surrounded by a background of true vacuum.

In a theory that does not allow for such non-trivial topologies (that is, where the vacuum manifold is simply connected) it is possible for scalar or fermion fields to interact with the soliton, called an embedded defect, and thus increase its stability.\footnote{there is a type of embedded string called a semilocal string that is completely stable, but it does not occur in Weinberg-Salam theory nor in the extension we are studying.} For example, in the Weinberg-Salam theory of Electroweak unification, embedded string and monopole solutions exist but they are unstable to small perturbations. This means that they are classically stable objects but they must decay quantum mechanically very soon after the phase transition in which they are formed. If this were all there were to say, they would be mere mathematical curiosities uninteresting to physicists. However, many authors for example Watkins and Vachaspati \cite{VW} have shown that by extending the standard model to include extra scalar fields, the stability of the Electroweak $Z$ string may be improved. Now that there is a solid reason to make this type of extension, as we have discussed, the extra fields appear naturally in noncommutative geometry with neutrino mixing, it is appropriate to consider their ideas and to apply them in our context.

\subsubsection{Why do they form?}

Consider the universe in the new phase. The order parameter cannot be isotropic across distances greater than that limited by causality: the field takes on different directions in different regions at random and it cannot be aligned at points separated by distances greater than light could have travelled since the beginning of the phase transition. Going back to the analogy with the ferromagnet, domains of a given spin orientation form after the symmetry is broken, and defects clearly occur at locations where the direction changes (if the spins change from up to down, at some point in between, they go to 0) and so topological defects are expected to proliferate near domain boundaries. The size of these boundaries is roughly uniform, and is determined by the speed of sound instead of the speed of light. The more quickly the phase transition takes place, the greater the number of defects form, as there is less time for communication between regions. This is why we observe more defects in a quenched specimen than an annealed one.\footnote{quenching can improve hardness, which is why the steel is quenched to make a Samurai sword.}

Due to this experience in the laboratory, one realises that if topological defects can form, they will form and this fact should be true in cosmology. (Of course, the analogy between topological defects formed in condensed matter physics experiments and their cosmological counterparts is not exact due to differences related to friction, relativistic effects, gravity, causality governed by light speed instead of speed of sound, knowledge of initial conditions and so on, but it can still be a useful one to make.) And, many grand unified theories predict that the solitonic solutions to the equations of motion in those theories must have formed in nature. Topological defects in cosmology can trap enormous amounts of energy because their cores contain old phase matter. Recall that in the symmetric phase the field is at the top of the potential hill and therefore the energy per unit volume will be much greater than that in the new phase. So, if they would not dominate the universe, they should at least have been detected by telescopes and show their signature in the cosmic microwave background data. As mentioned in the introduction, their absence is explained mainly by the fact that as the universe expands, distances between points increases and the defects could be already so far spread out that not even one of them occurs within our Hubble radius. Also, topological defects with opposite topological charge can annihilate eachother, by unwinding one another, and some cosmic strings can ``radiate away'' by emitting gravitational waves.

\subsubsection{How is torsion sourced by line defects?}

Cosmic strings and indeed embedded strings are line defects. They are analogous to the line dislocations and line disinclinations that are associated to broken translational and rotational symmetry respectively. The latter is a line source of curvature which can be pictured as a removal or addition of a wedge. The former is an edge or screw dislocation which are a source of torsion. Consider a particle in the material that transcribes a path around the dislocation. The result of this parallel transport is characterised by torsion whereas in the case of a disinclination, the result of a parallel transport around the defect is characterised by curvature. Therefore, an Electroweak string as a static configuration can be a source of non-propagating torsion.

\subsection{Electroweak $Z$ string solution}

Replacing the Electroweak gauge fields, and recalling the spectral Lagrangian (from Connes'
spectral action formula) with space-time curvature $R$ still set to 0:

\begin{equation*}
( \partial_{\mu} - ieA_{\mu})^{\ast} \phi^{\ast} (\partial^{\mu} -ieA^{\mu}) \phi
-\frac{1}{4}F_{\mu \nu}F^{\mu \nu} -\frac{\lambda}{4}(\vert \phi \vert^2 - \eta^2)^2
\end{equation*}

and the equation of motion for the scalar field $\phi^{\ast}$:

\begin{equation*}
(\partial_{\mu} + ieA_{\mu})(\partial^{\mu} -ieA^{\mu})\phi - \frac{\lambda}{2} \phi(\vert \phi \vert^2 - \eta^2)=0
\end{equation*}
or,  using the Lorentz gauge $\partial_{\mu}A^{\mu}=0$,

\begin{equation*}   \label{phi equation}
\square \phi  + \phi( e^2 A^2 - \frac{\lambda}{2}(\vert \phi \vert^2 - \eta^2)) = 0
\end{equation*}

where $\square$ is the d'Alembertian operator.

The equation of motion for the degrees of freedom in $D_F$, for example $t$, are found similarly,
but using the Euclidean Euler-Lagrange equations and hence there is a sign difference compared with
the above equations of motion. 

\begin{equation}     \label{string eom}
-4 \square t + t(e^2 A^2 + \vert t \vert^2 + \vert s \vert^2 + \vert p \vert^2 - 1)=0
\end{equation}

where the Higgs potential $-\frac{\lambda}{4}(\vert \phi \vert^2 - \eta^2)^2$ has been replaced by
($\vert t \vert^2 + \vert s \vert^2 + \vert p \vert^2 -1)^2$. Normally $\eta$ is set to 1. When
$p=0$ then $t$ and $s$ are just the usual Higgs doublet and the equation of motion is no different
from \ref{phi equation} except for the signature change.

In the Weinberg-Salam theory, the Ansatz for the Electroweak $Z$ string in polar coordinates is
$\phi(r,\theta) = f(r) e^{Y \theta} \phi_0$ where $Y$ is a (broken)\footnote{choice made in
\cite{KL} to minimise the magnetic energy} generator of the symmetry group of the action
in whose Lie algebra the gauge field $A$ takes values. (It has the same configuration as the Nielson-Olesen string, but it is not the same string as the latter is a topological defect in theories where the vacuum manifold is not simply connected). (The gauge field $A$ is part of the Ansatz
but since its equation of motion is not given above, it is not given here). \footnote{Kibble and
Lepora impose a condition so that there is no flux in the $r$ directions}. We can set up boundary
conditions so that $f(r) \rightarrow 1$ as $r \rightarrow \infty$ and we understand the statement
of \cite{KL} that the boundary conditions of the Electroweak strings define geodesic
paths on the vacuum manifold because the latter are given by $\gamma_Y = e^{Yt}$ where $t$ is some
real parameter. This discussion is developed in the final subsection.

\subsection{Stability of Electroweak strings}

Many authors have considered adding scalar fields to extend the Weinberg-Salam theory and carried out an analysis to determine whether their presence can increase the stability of the Electroweak $Z$ string. There are two strings in the Electroweak model, the $W$ and the $Z$. The former is unstable for all parameters while $Z$ is stable for a range of values, which in calculations done so far only include unphysical values. We will only discuss the $Z$ string. In  \cite{VW}, Watkins and Vachaspati add a singlet scalar field $\chi$ with global $U(1)$ charge and describe a ``bag'' mechanism as follows. The Higgs gives $\chi$ a mass by the usual Higgs mechanism, but the back-reaction of $\chi$ on the Higgs is to try to prevent it from acquiring its vev. In other words, the scalar would rather live in a region where Higgs vanishes since the mass of $\chi$ is zero wherever the Higgs is zero. The scalar likes to accumulate on the string and tends to maintain the string configuration. The string is a bag in which the scalar prefers to sit in and, hence, hold together.

In the current theory, the extra fields are not singlet scalars with a global $U(1)$ charge, but as described in \ref{field content}, $p$ is a complex scalar singlet with local $U(1)$ charge, and $(j,l)$ is a complex scalar doublet with weak isospin charge. The action terms in the leptoquarks (case $(D)$ in the notation of the last section) have weak isospin and colour charge.

Recall that $H$ and $G$ are non-zero in the false vacuum where $M=0$, and are 0 in the true vacuum where $M \neq 0$. Also recall the equations of motion for $H$ and $M$ calculated in \ref{field content}. Without giving too many details already published, we recall that in \cite{tsm} it was shown that $G$ and $M$ are mutually orthogonal complementary diagonal projections. (The vacuum solutions in either signature are that $D_F$ is a partial isometry and if we assume that $M$, $G$ are diagonalisable, - otherwise they would not have any physical meaning as a mass matrix - and working in mass eigenstates, then we can take the solutions to be diagonal projections). \emph{This means that the equations of motion for $G$ and $M$ or $H$ and $M$ precisely embody the bag phenomenon.}

Watkins and Vachaspati describe both a condensate and a bound state. A bound state consisting of the string and the $p$ particles can form as the bag, whereas a condensate would be a string in a background of $p$ particles. The bound state improves stability for a much larger range of parameters than the condensate. Their conclusion was that the bound state definitely improves the stability of the $Z$ string, but they could not find a stable solution within physical bounds (the stable solutions had too high a value for Weinberg angle). But they suggest a more extensive exploration of parameter space and an increase of the charge on the string. Since the fields involved in our context are to the order of fermion masses, it may be that these numbers improve things even more. For dynamical $H$ and $G$, we have a situation corresponding to this bound state, and where the leptoquarks are non-dynamical, we have the condensate.

In stability calculations, the starting point is to calculation the second differential of the potential with respect to the field. This is the same as the coefficient in the mass term and it is a significant term in $dE$ the change in energy stored in the defect core as the string solution is perturbed. We saw earlier that $\vt p \vv$ gave a positive contribution to the energy change by flattening off the hump, that is, if $p=0$, as in the standard model, the extremum at the symmetric phase is a maximum and any perturbations about that are obviously unstable. Whereas the presence of $p$ changes that. It is clear that $p \neq 0$ raises the energy in the string core since it tries to make the Higgs mass go positive. We conclude that the presence of $p$ significantly improves the stability of the Electroweak string. We can of course repeat the same argument for case $(D)$ and indeed $(D2)$, in which the physical parameters involve the degrees of freedom in $G$, which are related to the fermion masses. To determine exactly how significant these findings are, and to write a quantitative statement about the stability, which depends crucially on the values of the physical parameters involved, some numerical analysis would be required including solving a Schrodinger equation.

\subsection{The ``Higgs force'' and torsion gravity}  \label{torsion}

Connes creates a picture of two identical space-time sheets, one labelled left and the other
right. The geodesic distance between the two sheets is given by the metric (the
Dirac operator) on the noncommutative internal space. (See for example \cite{forces} and \cite{gravity and matter}). When this metric is flat, the sheets are a constant distance apart. There is a connection one-form describing parallel transport between the two sheets. As the metric is flat, it is a flat connection, its curvature is zero. The `flat' Dirac operator of internal space is then `fluctuated' by the noncommutative algebra, (this is called an `internal fluctuation') and by the equivalence principle, the
condition that the curvature is zero is relaxed, and it takes on the role of the Higgs field.
In this way, the Higgs scalar degrees of freedom are picked up. The Higgs
is promoted to a dynamical variable with non-zero curvature. So, now the metric is no longer
flat and the geodesic distance between the two space-time sheets is determined by  the eigenvalues of the fluctuated Dirac operator, which is of course, the fermion mass matrix. In the standard model, the internal space does not consist only of two points left and right, but models the whole of charge space. The mass matrix has a non-degenerate set of eigenvalues, as for each fermion there is a different distance between left and right.

The finite spectral action, (the Higgs potential \ref{fsa}), in which when varied with respect to all degrees of freedom in $D_F$, one gets a fully degenerate set of fermion masses, may be written as:

\begin{displaymath}
 S_{V(Higgs)} = \textrm{Tr}(MM^{\ast} -I)^2
\end{displaymath}

whereas if the equations of motion had a solution such that all the fermion masses were different, it would take a form of the type:

\begin{displaymath}  \label{G field}
 S_{curvature} = \textrm{Tr}(MM^{\ast} + GG^{\ast}-I)^2.
\end{displaymath}

In other words, there would be additional mixing terms between a set of new fields and the Higgs. Inspecting the derivation of the action in the spectral action principle, we may interpret this term as the \emph{curvature term for internal space}. So, to get a set of different masses, we need extra curvature giving rise to extra Higgs force via the equivalence principle. The action \ref{M and G action}  indeed provides this type of additional terms,  but $G$ is only non-zero in the vacuum of $M$ if the leptoquarks are non-propagating, that is, if they do not possess internal Einstein's field equations of their own.

Since the action is asking for an extra term, we are led to twist the Dirac operator one more time and use the Weintzenb\"ock equation such that the sought for additional curvature term may appear but we do not wish to add another $G$ field as in \ref{G field} in an ad hoc manner. At least the leptoquarks have a ra\^ison d'\^etre. We explain below why the extra term could just as well be due to torsion gravity instead of gravity from Riemannian geometry so that the new internal space gravity force is more like torsion gravity than curvature gravity. Arguably, it makes sense to interpret these extra gravity-making terms as torsion rather than curvature because: 

\begin{enumerate}
 \item $G$ satisfies a definition of an internal space torsion-type parallel transport suggested below, 
 \item the terms are intimately connected with the Electroweak string solutions, which are analogous to line defects in laboratory experiments that can source torsion, and
 \item to include torsion in the formalism is another natural extension, not an ``add-on.'' It is the removal of the constraint that unnaturally forces internal space torsion to be zero. 
\end{enumerate}

The commutative spectral triple is a Riemannian geometry, which means that the connection is defined to have zero antisymmetric part or torsion. The phrase Riemannian geometry is often taken to be synonymous with general relativity and the Higgs force is an analogy of gravity. A Riemann-Cartan geometry is one in which the connection is allowed to have a non-zero antisymmetric part and in this way it is a generalisation of general relativity. The gravitational field strength as usual, is described by the connection and it has not only a curvature 2-form but also a torsion 2-form. Torsion happens when we parallel transport two infinitesimal vectors in the tangent space at a given point along one another and the resulting parallelogram does not close. The torsion tensor is a measure of the non-closure of the resulting parallelogram. (This contrasts the notion of curvature, which characterises the difference between a vector before and after it is parallel transported around a loop on a curved manifold.)  In ``torsion gravity'' theories, the source of gravity is torsion instead of curvature. Unlike in torsion gravity theories, instead of intrinsic spin as the source of torsion, we have, if it be stable, the Electroweak string. So, by including torsion in the 0-dimensional spectral triple of the model we are studying, we are not making an unnatural addition, but simply removing the ad hoc constraint that the torsion be zero. The article \cite{Wodzicki} shows that the \emph{space-time} torsion in noncommutative geometry does not have the status of dynamical field as its kinetic term disappears (wherein there is a product with the symmetric metric tensor and the antisymmetric torsion).

$M$ and $M^{\ast}$ are  parallel transports between left and right and $MM^{\ast}$ is a parallel transport transcribing a complete loop, from left to right and back to left. It appears in the internal space curvature term identified in the action. $H$ is a map from right to right, and after a Wick rotation (as partially described in \ref{field content}), $G$ and $G^t$ are maps between left and left or right and right. So $H$ and $G$ are parallel transports \emph{in} the tangent space and the action terms in $G^{\ast}G$ should be a measure of torsion.

\subsection{Vacuum manifold metric and string energetics}

In \cite{KL} it was found that the string solutions define circular paths (geodesics) on the vacuum manifold.\footnote{In \cite{KL} sphalerons (Electroweak monopoles) are also studied.} We have illustrated this point earlier in this section. For this reason, Kibble and Lepora expected to find that the spectrum and properties of the strings should correspond to the geometry of the vacuum manifold. 
(The metric on the vacuum manifold is given by the Euclidean inner product at each tangent space.) So, the boundary conditions, which specify the solutions themselves, correspond to the geodesics with respect to that metric. They found two inequivalent metrics, one corresponding to each of the gauge and scalar sectors.  

What we find is that since there is a different metric on the internal space for each fermion (see the first paragraph of \ref{torsion}), which in turn determines the metric on the vacuum manifold, there must be a set of geodesic paths with respect to each of those metrics. (The three-sphere $\vt \phi_1 \vv + \vt \phi_2 \vv =1$ becomes  $\vt a \vv + \vt b \vv = 1 - \vt g \vv - \vt h \vv$ for the up quark and so on. To avoid repetition of these equations, we refer to \cite{tsm} for the corresponding metric pertaining to the other fermions.) For a given metrical structure, one has a corresponding curvature of the vacuum manifold to which the energy of the embedded vortex is associated. Given a stable string, these metrics and geodesics paths correspond to that string's boundary conditions. In other words, since there is a discrete set of masses, there is a discrete set of metrics on the vacuum manifold with respect to which geodesic paths on the vacuum manifold correspond to embedded string solutions in this extension to the Weinberg-Salam theory, whose stabilities are connected to the fermion masses as we have seen.

\textbf{Acknowledgements}

I would like to thank my teachers, Levon Pogosian and John W. Barrett. 
\\This research was supported by FCT (Fundaç\~ao para a Ci\^encias e a Tecnologia) and it was carried out under the affiliation given on the title page.


\begin{thebibliography}{999}

\bibitem[BES]{VW} T. Vachaspati, R. Watkins, \emph{Bound states can stabilize electroweak strings.} Phys. Lett. B 318 (1993) 163-168.
\bibitem[C]{Connes' book} A. Connes \emph{Noncommutative Geometry.}
Academic Press, London, (1994).
\bibitem[CW]{cw} S. Coleman, E. Weinberg, \emph{Radiative corrections as the origin of spontaneous symmetry breaking.} Phys. Rev. D 7. 1888-1910 (1973).
\bibitem[DJ]{dj} L. Dolan, R. Jackiw, \emph{Symmetry behaviour at finite temperature.} Phys. Rev. D 9. 3320-3341 (1974).
\bibitem[EVG]{KL} F. Lepora, T.W.B. Kibble, \emph{Electroweak vacuum geometry} J. High Energy Phys. JHEP04 (1999) 027.
\bibitem[FCG]{forces} T. Sch\"ucker, \emph{Forces from Connes' geometry.} ArXiv:hep-th/0111236.
Lect. Notes Phys. 659:285-350 (2005).
 \bibitem[GM]{gravity and matter} A. Connes, \emph{Gravity coupled with matter and the foundation of
non-commutative geometry.} Comm.  Math. Phys. Vol.182 (1996), N.1, 155-176.
\bibitem[GSN]{gsn} A.H. Chamseddine, A. Connes, M. Marcolli, \emph{Gravity and the standard model with neutrino mixing.} Arxiv: hep-th/061024 (2006).
\bibitem[INS]{Landi} G. Landi, \emph{An introduction to noncommutative spaces and their topology.}
Arxiv: hep-th/9701078 (1997).
\bibitem[KTS]{kibble strings} T.W.B. Kibble, \emph{Topology of cosmic domains and strings.} J. Phys. A 9: Math. Gen. 1387 (1976).
\bibitem[L]{Landi's book} G. Landi, \emph{An Introduction to
Noncommutative spaces and their geometries.} Springer-Verlag (1998).
\bibitem[LNG]{lng} J.W.Barrett, \emph{A Lorentzian version of the non-commutative geometry of the standard model of particle physics.} J. Math. Phys. 48 012303 (2007).
\bibitem[NGL]{leptoquarks}
M. Paschke, F. Scheck, A. Sitarz, \emph{Can (noncommutative) geometry accommodate leptoquarks?}
Arxiv: hep-th/9709009. Phys. Rev. D59:035003 (1999).
\bibitem[NGR]{non-com and reality} A.Connes, \emph{Non-commutative geometry and reality.} J. Math.
Phys. Vol.36 (1995), 6194.
\bibitem[NOA]{noa} C.A. Stephan, \emph{Almost-commutative geometry, massive neutrinos and the orientability axiom in KO-dimension 6.} Arxiv: hep-th/0610097 (2006).
\bibitem[NSM]{ncg and sm} T. Sch\"ucker, \emph{Noncommutative geometry
and the standard model.} Arxiv: hep-th/0409077. Int. J. Mod. Phys. A20:2471-2480 (2005).
\bibitem[OCI]{Schucker} B. Iochum, T. Sch\"ucker, C. Stephan, \emph{On a classification of
irreducible almost commutative geometries.} J. Math. Phys. Vol.45 (2004), 5003-5041.
\bibitem[R]{rivers} R. Rivers, \emph{Path Integral Methods in Quantum Field Theory.} Cambridge University Press.
\bibitem[RCS]{RCS} M. Sakellariadou, \emph{The revival of cosmic strings.} Ann. Phys. 15, No. 4-5, (2006) 264-276.
\bibitem[SAP]{sap} A.H. Chamseddine, A. Connes, \emph{The spectral action principle.} Comm. Math.
Phys. Vol.186 (1997), N.3, 731-750.
\bibitem[SMV]{smv} J.W. Barrett, R.A. Martins, \emph{Non-commutative geometry and the standard
model vacuum.} Arxiv: hep-th/0601192, J. Math. Phys. 47 (2006) 052305.
\bibitem[SNM]{snm} A. Connes, \emph{Noncommutative geometry and the standard model with neutrino mixing.} J. High Energy Phys. JHEP11 (2006) 081.
\bibitem[TSM]{tsm} R.A. Martins,  \emph{Noncommutative geometry, topology and the standard model vacuum.} Arxiv: hep-th/0609140, J. Math. Phys. 47 (2006) 113507.
\bibitem[VL]{vachaspati lectures} T. Vachaspati, \emph{Lectures on cosmic topological defects.} Arxiv: hep-ph/0101270 (2001).
\bibitem[VS]{vilenkin shellard} A. Vilenkin, E.P.S. Shellard, \emph{Cosmic strings and other topological defects.} Cambridge Monographs (1993).
\bibitem[WR]{Wodzicki} W. Kalau, M. Walze, \emph{Gravity, non-commutative geometry and the Wodzicki Residue.} Arxiv: gr-qc/9312031, J. Geom. Phys. 16 (1995) 327-344.
 \end{thebibliography}
\end{document}